\documentclass[12pt,a4paper]{article}
\usepackage{graphicx}

\def\simgt{\stackrel{>}{{}_\sim}}
\def\simlt{\stackrel{<}{{}_\sim}}

\begin{document}
\pagestyle{empty}

{\Large\bf A faint type of supernova from a white dwarf with a helium-rich companion}
\vspace{1cm}
\begin{centering}

H.~B.~Perets,\\
{\scriptsize Department of Particle Physics and 
Astrophysics, Faculty of Physics,} 
{\scriptsize The Weizmann Institute of Science,
Rehovot 76100, Israel;}\\
{\scriptsize and CfA Fellow, Harvard-Smithsonian Center for Astrophysics, 60 Garden Street, Cambridge, MA 02138, USA}\\ 
A. Gal-Yam,\\
{\scriptsize Department of Particle Physics and 
Astrophysics, Faculty of Physics,} 
{\scriptsize The Weizmann Institute of Science,
Rehovot 76100, Israel}\\
P. A. Mazzali,\\
{\scriptsize Max-Planck Institut f\"{u}r Astrophysik, Karl-Schwarzschild-Str. 
  1, 85748 Garching, Germany;}\\
{\scriptsize Scuola Normale Superiore, Piazza Cavalieri 7, 56127 Pisa, Italy;}\\
{\scriptsize and INAF -- Oss. Astron. Padova, vicolo dell'Osservatorio, 5, 35122 Padova, Italy}\\
D. Arnett, \\
{\scriptsize Steward Observatory, University of Arizona, 933 
North Cherry Avenue, Tucson, AZ 85721, USA}\\
D.~Kagan,\\
{\scriptsize Department of Astronomy, University of Texas at Austin, Austin, TX 78712, USA}\\
A.~V.~Filippenko,\\
{\scriptsize Department of Astronomy, University of California, Berkeley, CA 94720-3411, USA}\\
W.~Li,\\
{\scriptsize Department of Astronomy, University of California, Berkeley, CA 94720-3411, USA}\\
I.~Arcavi,\\
{\scriptsize Department of Particle Physics and 
Astrophysics, Faculty of Physics,}
{\scriptsize The Weizmann Institute of Science,
Rehovot 76100, Israel}\\
S.~B.~Cenko,\\
{\scriptsize Department of Astronomy, University of California, Berkeley, CA 94720-3411, USA}\\
D.~B. Fox,\\
{\scriptsize Dept.~of Astronomy and Astrophysics, Pennsylvania 
State University, University Park, PA 16802, USA}\\
D.~C.~Leonard,\\
{\scriptsize Department of Astronomy, San Diego State University, San Diego,California 92182, USA}\\
D.-S.~Moon,\\
{\scriptsize Department of Astronomy and Astrophysics, University of Toronto, 50 St.~George Street, Toronto, ON M5S 3H4, Canada}\\
D.~J.~Sand,\\
{\scriptsize Harvard-Smithsonian Center for Astrophysics, 60 Garden Street, Cambridge, MA 02138, USA}\\ 
{\scriptsize and Las Cumbres Observatory Global Telescope Network, 6740 Cortona Dr., Suite 102, Goleta, CA 93117, USA}\\
A.~M.~Soderberg,\\
{\scriptsize Harvard-Smithsonian Center for Astrophysics, 60 Garden Street, Cambridge, MA 02138, USA}\\ 
J.~P.~Anderson,\\
{\scriptsize Departamento de Astronomía, Universidad de Chile, Camino El Observatorio 1515, Las Condes, Santiago, Casilla 36-D, Chile;}\\
{\scriptsize and Astrophysics Research Institute, Liverpool John Moores University, Twelve Quays House, Birkenhead CH41 1LD, UK}\\
P.~A.~James,\\
{\scriptsize Astrophysics Research Institute, Liverpool John Moores University, Twelve Quays House, Birkenhead CH41 1LD, UK}\\
R.~J.~Foley,\\
{\scriptsize Clay Fellow, Harvard-Smithsonian Center for Astrophysics, 60 Garden Street, Cambridge, MA 02138, USA}\\ 
M.~Ganeshalingam,\\
{\scriptsize Department of Astronomy, University of California, Berkeley, CA 94720-3411, USA}\\
E.~O.~Ofek,\\
{\scriptsize Department of Astronomy, 105-24, California 
Institute of Technology, Pasadena, CA 91125, USA}\\
L.~Bildsten,\\
{\scriptsize Kavli Institute for Theoretical Physics, Kohn Hall, University of California, Santa Barbara, CA 93106, USA}\\
{\scriptsize and Department of Physics, University of California, Santa Barbara, CA 93106, USA} \\
G.~Nelemans,\\
{\scriptsize Department of Astrophysics, Radboud University Nijmegen, P.O. Box 9010, NL-6500 GL, The Netherlands}\\
K.~J.~Shen,\\
{\scriptsize Kavli Institute for Theoretical Physics, Kohn Hall, University of California, Santa Barbara, CA 93106, USA}\\
N.~N.~Weinberg,\\
{\scriptsize Department of Astronomy, University of California, Berkeley, CA 94720-3411, USA}\\
B.~D.~Metzger,\\
{\scriptsize Department of Astronomy, University of California, Berkeley, CA 94720-3411, USA}\\
A.~L.~Piro,\\
 {\scriptsize Department of Astronomy, University of California, Berkeley, CA 94720-3411, USA}\\
E.~Quataert,\\
{\scriptsize Department of Astronomy, University of California, Berkeley, CA 94720-3411, USA}\\
M.~Kiewe,\\
{\scriptsize Department of Particle Physics and 
Astrophysics, Faculty of Physics,} 
{\scriptsize The Weizmann Institute of Science,
Rehovot 76100, Israel}\\
and D.~Poznanski,\\
{\scriptsize Department of Astronomy, University of California, Berkeley, CA 94720-3411, USA} \\
{\scriptsize and Lawrence Berkeley National Laboratory, 1 Cyclotron Road, Berkeley, CA 94720, USA}

\end{centering}

\date{\today}{}
\clearpage

{\bf Supernovae (SNe) are thought to arise from two different physical processes. 
The cores of massive, short-lived stars undergo gravitational core collapse and typically eject a few solar masses during their explosion. 
These are thought to appear as as type Ib/c and II SNe, and are associated with young stellar populations. 
A type Ia SN is thought to arise from the thermonuclear detonation of 
a white dwarf star composed mainly of carbon and oxygen, whose mass approaches the 
Chandrasekhar limit\cite{fil97,maz+07}. Such SNe are observed in both 
young and old stellar environments. Here we report our discovery of the 
faint type Ib SN 2005E in the halo of the nearby isolated galaxy, NGC 1032. 
 The lack of any trace of recent star formation near  the SN location (Fig. 1), and the very low derived ejected mass 
($\sim0.3$ M$_\odot$), argue strongly against a core-collapse 
origin for this event. Spectroscopic observations 
and the derived nucleosynthetic output show that the SN ejecta have 
high velocities and are dominated 
by helium-burning products, indicating that SN 2005E 
was neither a subluminous\cite{fil+92,li+03} nor a regular\cite{fil97} SN~Ia (Fig. 2).
We have therefore found a new type of stellar explosion, arising from a low-mass, old stellar system,
likely involving a binary with a primary white dwarf and a helium-rich secondary. The SN ejecta contain more calcium
than observed in any known type of SN and likely 
additional large amounts of radioactive $^{44}$Ti.
Such SNe may thus help resolve fundamental physical puzzles, 
extending from the composition of 
the primitive solar system and that of the oldest stars, to the Galactic 
production of positrons.}
\clearpage



We discovered a supernova (SN) explosion (SN 2005E; Fig. 1) on Jan. 13, 2005 (UT dates are used throughout this paper) shortly
after it occurred (it was not detected on Dec. 24, 2004). Follow-up
spectroscopy (Fig. 2) 
revealed strong lines of helium and calcium, 
indicating that it belongs to the previously identified
group of calcium-rich type Ib SNe\cite{fil+03}.  
The SN position is  $\sim22.9$ kpc (projected) from the centre and $\sim11.3$
kpc above the disk of its edge-on host galaxy, NGC 1032 (Fig. 1),
which is itself at a distance of 34 Mpc. 
NGC 1032 is an isolated galaxy\cite{pra+03} 
showing no signs of interaction, with the closest
small satellite galaxy found at a distance $>120$ kpc in projection.
Deep follow-up observations of the explosion site, 
sensitive to both ultraviolet light from hot young stars and emission lines
from ionized hydrogen gas, put strict limits on any local star-formation 
activity at or near the SN location (Fig. 1). In addition, a radio signature, 
expected from some core-collapse SNe, has not been observed 
(see Supplementary Information; SI, Section 2).

Our analysis of the spectra of SN 2005E indicates that it is 
similar to SNe Ib (Fig. 2 and SI, Section 3), 
showing lines of He but lacking either
hydrogen or the hallmark Si and S lines of SNe Ia in its photospheric spectra. 
The nebular spectrum of this event shows no emission from iron-group elements,
 which also characterize type Ia SNe (SI, Sections 3 and 4). 
Analysis of this spectrum indicates a total 
ejected mass of $\rm{M}_{ej}\approx0.275$ M$_{\odot}$, 
with a small fraction in radioactive nickel, 
consistent with the low luminosity 
of this event. 
Such low ejecta mass for a SN of any type has never before been
firmly established using nebular spectral analysis (SI, Fig. S3 and Section 5).
We also used the narrow, fast, and faint light curve (SI, Fig. S4 in Section 6)
 together with the measured ejecta velocity 
($\sim 11,000$ km s$^{-1}$) to infer the ejected mass 
(SI, Section 6). We use these data to find consistent 
results of $M_{\rm ej}\approx 0.3\pm0.1$ M$_{\odot}$, assuming that some of the 
mass is not accounted for by the nebular spectrum analysis (e.g., 
high-velocity He layers and some slowly moving, denser ejecta that are 
still hidden below the photosphere at that time).
Finally, SN 2005E exhibits a remarkable amount
of calcium in its ejecta, $0.135\,\rm{M}_{\odot}$ 
($\sim0.49$ of the total ejecta mass),  5--10 times
more than typical SNe of any variety, with a relative calcium fraction 
25--350 times higher than any reported values for other SNe   
(see Table 1 and SI, Section 7), while not showing evidence for sulfur (SI, Section 4). 

The remote position of SN 2005E in the outskirts (halo) of the galaxy, together
with the isolation of NGC 1032 and its classification as an S0/a galaxy
(in which the star-formation rate is very low\cite{ken98}), in addition to 
our limits on local star formation, point to
a SN progenitor from an old stellar population 
(see also SI, Section 2). In addition, the low ejected mass and nucleosynthetic 
output of SN 2005E are in stark contrast to those expected from
collapsing massive stars, whether formed locally or ejected from a distant 
location (SI, Sections 8--9). 

The low ejected mass is also inconsistent 
with those determined for SNe~Ia, restricted to a tight 
mass range of $\sim1$--$1.3\,\rm{M}_{\odot}$, 
regardless of their intrinsic luminosity (even the prototype faint SN 1991bg 
is found in this range)\cite{maz+07}. Furthermore, the light curve of SN 2005E
(see SI, Section 6) shows a different behavior than that of SNe~Ia,
declining much faster than even the most subluminous (SN 1991bg-like)  
events observed\cite{kas+08}.   
These properties, together with the observed He-rich spectra and inferred
composition, rule out SN 2005E as being either a regular or peculiar SN~Ia 
(see also discussion in the SI, Section 10, regarding the very subluminous 
SN 2008ha and other related peculiar SNe\cite{li+03,fol+09,val+09}).
Therefore, we conclude that SN 2005E 
is the first clearly identified example of a new, 
different type of SN explosion, arising from a He-rich, low-mass
progenitor. 

The spectroscopic signatures of SN 2005E are quite unusual, and allow
 one to identify additional similar events\cite{fil+03}. Arising from 
lower-mass progenitors, these events are likely to be found among both 
old and young stellar populations --- that is, we expect to find such peculiar 
SNe~Ib in both early- and late-type galaxies. Indeed, while the unusual 
location of SN 2005E triggered the current study, several other calcium-rich 
subluminous SNe~Ib/c similar to SN 2005E have been observed (SI,  Section 11). 
Of the group of eight subluminous calcium-rich  SNe~Ib/c identified (seven 
identified by us and an additional one described in Ref. \cite{kaw+09}), 
four are observed in old-population environments: SN 2005E presented here, 
as well as SNe 2000ds, 2005cz, and SN 2007ke, residing in elliptical galaxies. 
SN 2000ds has pre- and post-explosion {\it Hubble Space Telescope} images 
showing no evidence for either star-forming regions or massive 
stars\cite{mau+05} 
near its location. The host-galaxy distribution of the SNe in our sample 
(Fig. 3) is inconsistent with that of any core-collapse SN. 
No radio signatures have been observed either (see SI, Section 11). 
 Thus, all evidence suggests that a well-defined subset of SNe~Ib, all having Ca-rich spectra and faint peak magnitudes, 
comprise a distinct physical class of 
explosions coming from low-mass, old progenitors. This class includes all 
known type Ib/c events in confirmed elliptical 
galaxies\cite{van+05,hak+08} (SI,  Section 11). A different interpretation\cite{kaw+09}, 
invoking the core collapse of a massive progenitor, was suggested for one of these
events (SN 2005cz). It is difficult
to reconcile our observations and analysis of SN 2005E with such an interpretation, which is 
also inconsistent with the host-galaxy distribution of all of the other Ca-rich SNe~Ib in our sample. 

Calcium-rich SNe were theoretically predicted to arise from
 burning helium-rich material on a WD (e.g., a helium WD or a helium star 
accreting onto a CO WD), leading to the full disruption
of a sub-Chandrasekhar-mass WD\cite{woo+86,woo+94}. 
However, such models predicted
the production of SNe far more luminous (and $^{56}$Fe rich) than SN 2005E. 
Several theoretical models were suggested in the literature to possibly 
produce subluminous SNe, with low-mass and high-velocity ejecta 
in an old stellar population. These include the accretion-induced 
collapse (AIC) of a WD (e.g., Refs. \cite{nom+91} and \cite{met+09}), 
and the detonation of an accreted helium 
shell on a WD in a binary system (the ``.Ia'' model\cite{bil+07}).
These studies did not explore the burning of large
helium masses ($>0.1\,{\rm M}_\odot$), nor the production 
of calcium-rich ejecta. 
Multi-dimensional simulations of a detonation in accreted He layers\cite{liv+95}
showed (for low-mass white dwarfs; $M = 0.7\,{\rm M}_\odot$) a trend 
toward large Ca abundances and high Ca/S abundance ratio 
(a high ratio is inferred for SN 2005E; see SI, Section 4) 
and a light curve that was faster and dimmer 
than those of typical SNe~Ia (but still much more luminous than SN 2005E),
as well as a high production of $^{44}$Ti. 
It is possible that similar models, with less burning of C and O
to make S and Ni, may resemble SN 2005E. This gains additional support from 
our nucleosynthetic analysis (see SI, Section 12), showing that the unique 
composition of SN 2005E could be produced, in principle, 
as the product of He ignition. Further studies in these directions 
are in progress.

We conclude that SN 2005E appears to be the first observed manifestation of 
the helium detonation process.
This event most likely
occurred in an interacting double WD system with a helium WD mass donor. 
Additional characteristics  
of these explosions, including their old population origin, He-rich spectra,
subluminosity, and low ejected mass, are broadly consistent with  
the predictions of some theoretical models 
(.Ia\cite{bil+07}; AIC\cite{met+09}; helium detonation\cite{ibe+87}), variants of which 
may produce the appropriate conditions for such helium detonations. 
Alternatively, these explosions may require a totally new mechanism.

Our discovery has numerous
astrophysical implications. It seems highly likely that we 
identified explosions arising from very close WD-WD systems,  
the rates of which (SI, Section 11) might be useful for predicting 
the rates of WD-WD inspirals observable as gravitational wave sources.  
The unique nucleosynthetic production of large masses of calcium and 
radioactive $^{44}$Ti per explosion could solve 
puzzles related to the source of calcium (especially $^{44}$Ca) in the 
primitive solar system\cite{woo+73,tim+96} and
in old, metal-poor
halo stars\cite{lai+09}, and the enrichment patterns 
of the interstellar and intracluster medium\cite{dep+07}.
Production of most of the
Galactic $^{44}$Ti and its progeny, $^{44}$Ca, in a few
rare, prolific explosions, can also explain the origins of 
Galactic $^{44}$Ca given the null detection of $^{44}$Ti 
traces in most nearby SN remnants\cite{tim+96,the+06}.

Finally, inverse $\beta$ decay of $^{44}$Ti 
may significantly contribute to the Galactic production 
of positrons\cite{cha+93}.
Assuming our estimated rates ($\sim10\%$ of the SN~Ia
rate; SI, Section 11) and our $^{44}$Ti yield (0.014--0.14 M$_\odot$; 
SI, Section 12), Galactic SNe of the type 
we describe here will provide a significant contribution
to the Galactic bulge component of the positron annihilation line, at least
comparable to that of SNe~Ia. In fact, within the current 
uncertainties on the $^{44}$Ti yield and SN rates, these
events may come within a factor of a few of producing all of the 
observed positrons\cite{kno+05}.


\clearpage

~\\
We would like to thank P. Podsiadlowski, E. Nakar, D. Maoz, and the 
referees for helpful comments. We acknowledge observations with
the Liverpool Telescope, and various telescopes at the Lick, Palomar,
and Keck Observatories.
We are grateful to the staffs of these observatories, as well as to
the institutions, agencies, and companies funding these facilities.
This research has also made use of the NASA/IPAC
Extragalactic Database (NED).
H.B.P. acknowledges the ISF/FIRST and Ilan Ramon-Fulbright Fellowships.    
The collaborative work of A.G. and P.M. is supported by a 
Weizmann-Minerva grant. A.G. acknowledges further support by the Israeli 
Science Foundation, an EU Seventh Framework Programme Marie Curie IRG 
Fellowship, the Benoziyo Center for Astrophysics, and the Peter and Patricia 
Gruber Awards. A.V.F. is grateful for the support of the US National
Science Foundation, the US Department of Energy, Gary and Cynthia
Bengier, the Richard and Rhoda Goldman Fund, the Sylvia \& Jim Katzman
Foundation, and the TABASGO Foundation.

~\\ H.B.P. led the project, performed the calculations related to
hyper-velocity stars, examined other putative SN 2005E-like events,
collected and analyzed archival data concerning SN properties and
their hosts, and wrote the manuscript. A.G.-Y. is the PI of the CCCP
program and initiated the project, collected and analyzed photometric
and spectroscopic data, coordinated further observational and
theoretical work, and managed the project. P.A.M. conducted the
nebular spectral analysis and its interpretation, and determined the
elemental abundances in the ejecta. D.A. determined that the measured
composition requires He burning and performed nucleosynthesis
calculations to confirm this. D.K. investigated local star-formation
tracers at the location of SN 2005E. A.V.F. and W.L. contributed
spectroscopic and photometric observations and reductions of SN 2005E
and of similar Ca-rich objects, a class they originally identified,
and provided most of the data on SN host galaxies. A.V.F. also
carefully edited the paper. I.A. analyzed the CCCP photometry of SN
2005E and cross-calibrated it with other data. S.B.C., D.B.F., D.C.L.,
D.-S.M., D.J.S., and A.M.S. are members of the CCCP and contributed to
initial observations of SN 2005E. J.P.A. and P.A.J. obtained and
analyzed narrow-band images of NGC 1032 and the location of SN 2005E.
R.J.F. and M.G. contributed to spectroscopic observations and
reductions. E.O.O. obtained deep photometric observations of the
location of SN 2005E. L.B., G.N. K.J.S., and N.N.W. investigated the
relation of SN 2005E to .Ia models and contributed to the
text. B.D.M., A.L.P., and E.Q. investigated the relation of SN 2005E to
AIC models and contributed to the text. M.K. performed custom
reductions of CCCP spectra. D.P. carried out synthetic photometry
analysis.

%

~\\
The authors declare no competing financial interests.\\

\noindent Correspondence should be addressed to A. Gal-Yam 
(avishay.gal-yam@weizmann.ac.il)
and H. B. Perets (hperets@cfa.harvard.edu).\\

\noindent Supplementary Information accompanies the paper on www.nature.com/nature\\

\noindent

{%

\centerline{\includegraphics[width=4.7in,angle=0]{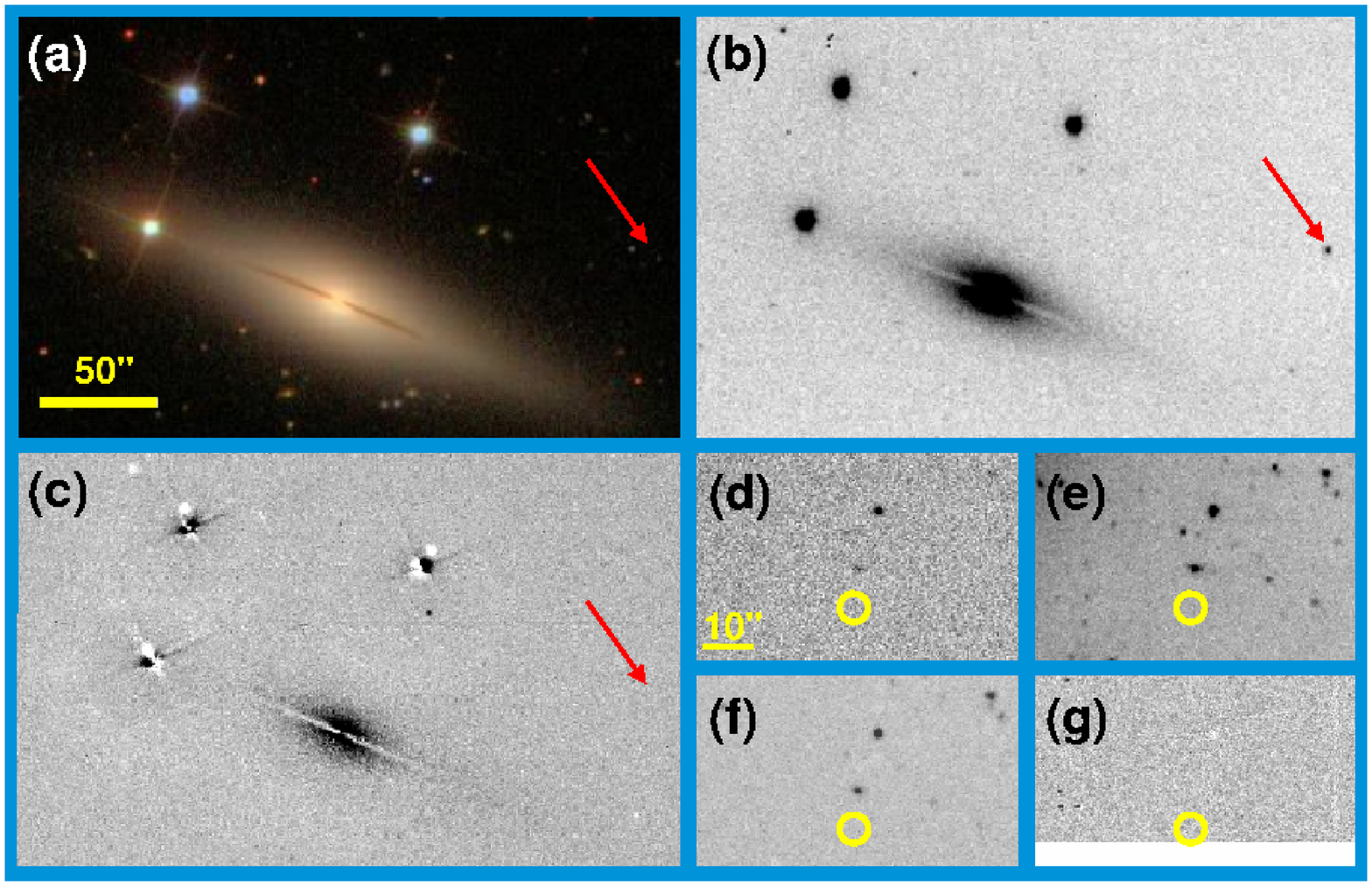}}
{\bf Figure 1}:
The environment of SN 2005E (technical details about the
observations can be found in SI, Section 1). {\bf (a)} NGC 1032, the host galaxy 
of SN 2005E, as observed by the Sloan Digital Sky Survey (SDSS), prior to the 
SN explosion. The galaxy is an isolated, edge-on, early-type spiral galaxy, showing no signs
of star-formation activity, warping, or interaction. Its luminosity is dominated by the cumulative 
contribution of a multitude of low-mass old stars (yellow light in this image). Panels
{\bf (a)-(c)} are $275'' \times 175''$; a scale bar is provided, north is up, and east is to the left.       
{\bf (b)} The LOSS\cite{fil+01} discovery of SN 2005E on Jan. 13, 2005 
(shown in negative). Note the remote location
of the SN (marked with a red arrow) 
with respect to its host, 22.9 kpc (projected) from the galaxy nucleus and
11.3 kpc above the disk, whose edge-on orientation is well determined (panel {\bf (a)}).
{\bf (c)} An image of NGC 1032 in the light of the 
H$\alpha$ emission line, emitted by interstellar gas ionized by 
ultraviolet (UV) radiation, and a good tracer of recent star formation. 
There are no traces of recent star-formation activity (usually appearing as 
irregular, compact emission sources) near the SN location or anywhere else
 in the host. 
Panels {\bf (d)-(g)} are $64'' \times 36''$; a scale bar is provided. 
{\bf (d)} Zoom-in on the location of SN 2005E in pre-explosion 
SDSS $r$-band images. No source is detected near the SN location, 
marked with a yellow circle 
(radius $3''$; the astrometric 
uncertainty in the SN location is $<0.5''$). 
The SDSS catalog does not list any 
objects near that position (e.g., putative faint dwarf satellites of NGC 1032),
 down to a typical limit of $r=22.5$ mag. {\bf (e-f)}
Deeper photometry of the SN location. A red image 
 is shown in panel {\bf (e)}, while
a UV ($u$-band) image is shown in panel {\bf (f)}.
 At the distance of NGC 1032, the point-source upper limits we find,
 $M_r<-7.5(-6.9)$  and $M_{u'}<-8.1(-7.1)$ mag at $3(2)\sigma$, 
respectively, indicate that we would have detected faint star-forming 
galaxies or star-forming regions  at the SN location, or indeed 
even individual massive 
red supergiant or luminous blue supergiant stars. 
{\bf (g)} Zoom-in on the location of SN 2005E in H$\alpha$ light 
(see panel {\bf (c)} for details). 
No trace of star-formation activity is seen near the SN location.}


\clearpage

\noindent

{%
\centerline{\includegraphics[width=4in,angle=0]{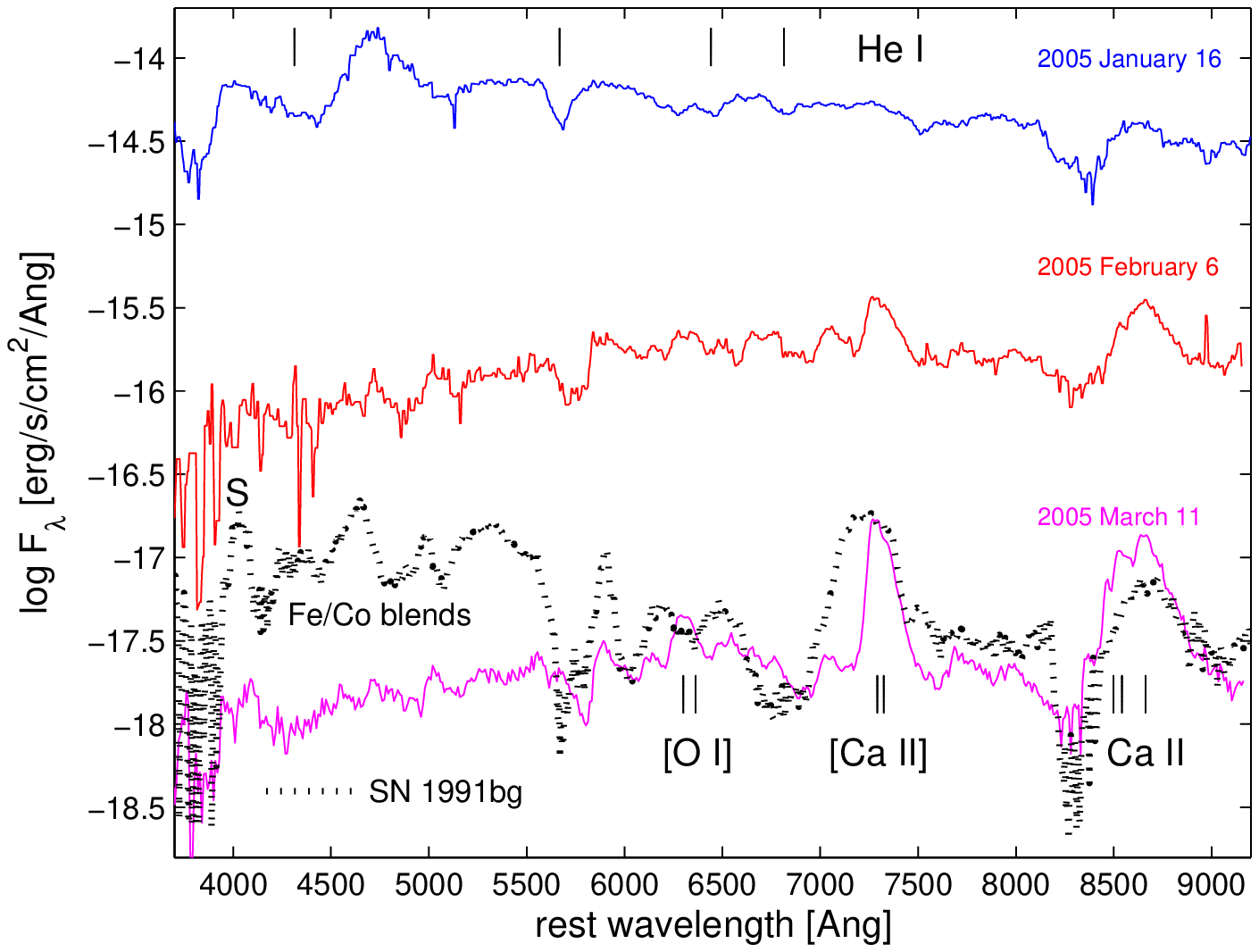}}
\centerline{\includegraphics[width=4in,angle=0]{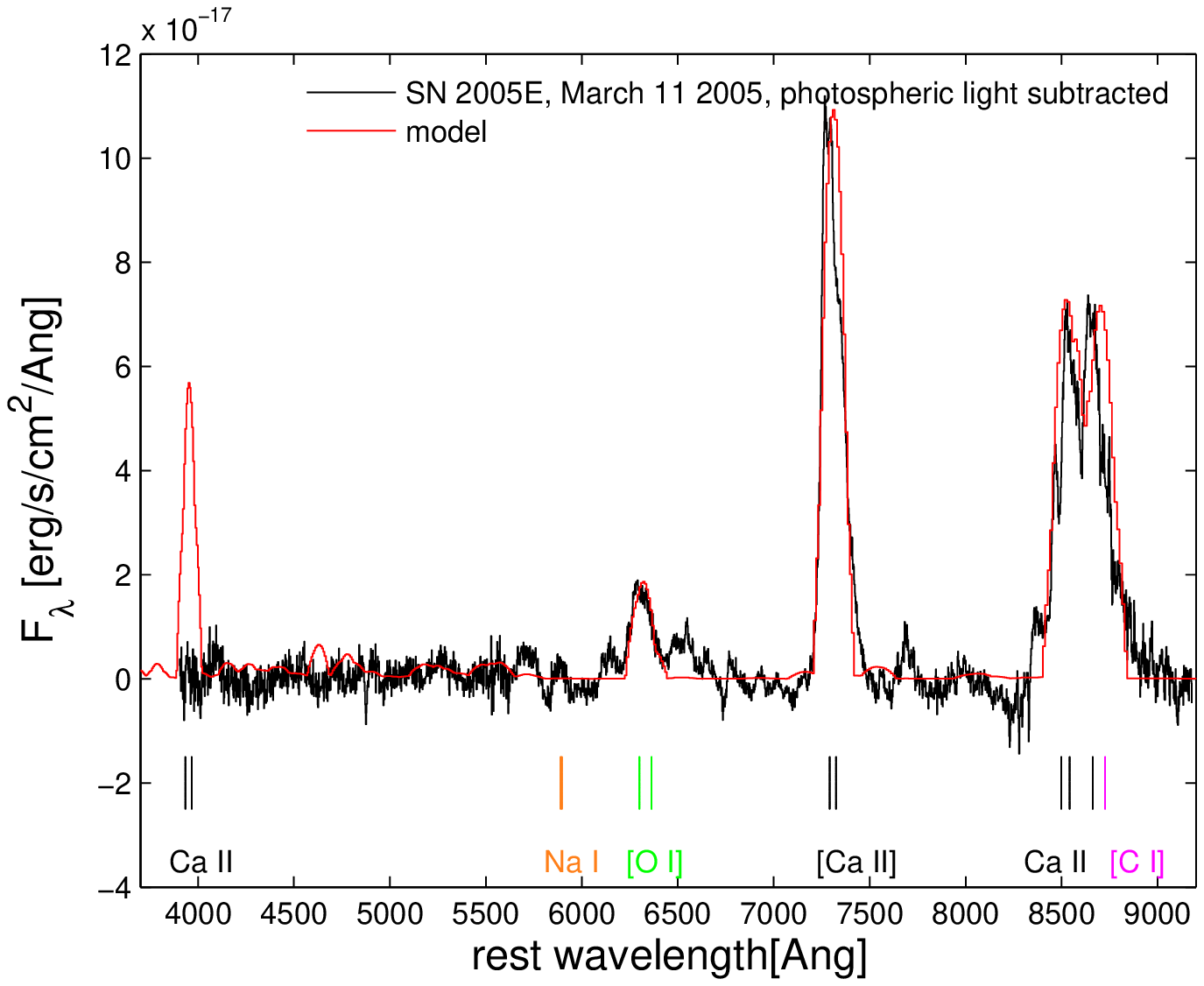}}

{\bf Figure 2}:
The mass and composition 
of the SN 2005E ejecta 
(technical details of observations and additional references can 
be found in SI, Section 1). 
Upper panel: Photospheric spectra of SN 2005E. 
The top spectrum is obviously photospheric and shows absorption lines of the
 He~I series (marked with black ticks after application of an 11,000 km s$^{-1}$ blueshift, at the top). 
Nebular lines of intermediate-mass elements, most notably calcium,
 begin to emerge in the middle spectrum. Calcium dominates the latest nebular 
spectrum at the bottom, and nebular oxygen is visible as well. 
Also note that the typical Si lines of SNe~Ia are absent in all spectra, while
the nebular spectrum of SN 2005E clearly rules out a type Ia identification
(comparison with the underluminous SN 1991bg is shown; 
note the lack of the typical iron-group line blends in the blue side). 
The derived line velocities are consistent with SN 2005E exploding within its 
putative host galaxy, NGC 1032.
Bottom: The nebular spectrum of SN 2005E compared with a model fit.
From the fit we can derive
elemental abundances and masses in the ejecta of SN 2005E. We find masses
of 0.1, 0.037, 0.135, and 0.003 $\rm{M}_\odot$ for carbon, oxygen, calcium, and 
radioactive nickel, respectively. Both the low total ejected mass of $\sim0.275$ M$_{\odot}$ and the relative
abundances are unique among previously studied events. 
The lack of prominent C/O-burning products 
such as S and Fe (typically seen in SNe~Ia; SI, Section 4) 
argues against a C/O WD origin.}

\clearpage
\noindent

\centerline{\includegraphics[width=4.5in,angle=0]{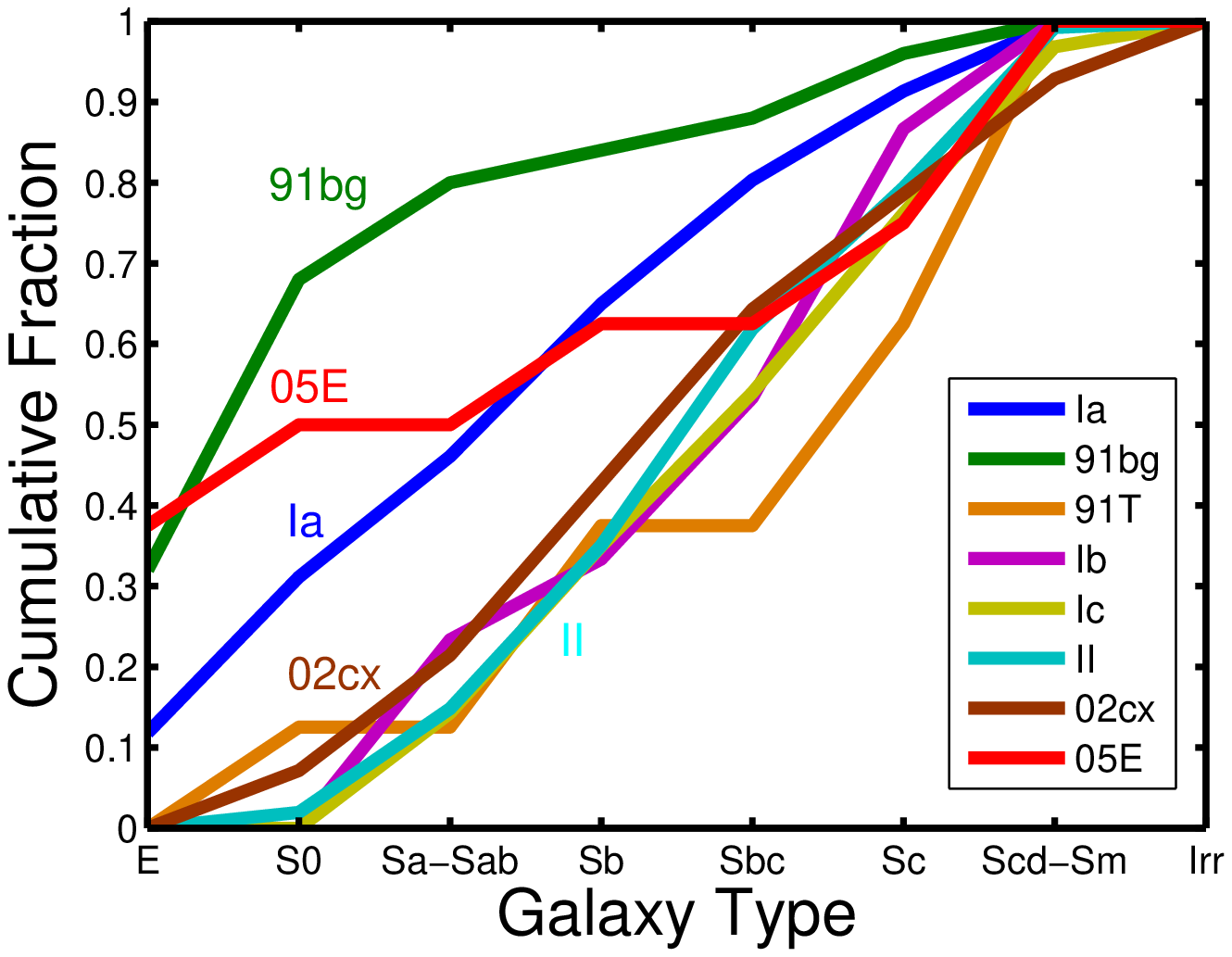}}
{\bf Figure 3}:
  The cumulative distribution of host galaxies of SNe from the KAIT SN 
survey. We corrected the classification of a few SN~Ib/c hosts
using higher-quality observations from the Palomar 60-inch telescope
(SN 2005ar, 2006ab, and 2006lc were found to be hosted by spiral galaxies 
rather than elliptical galaxies).  
After correcting the classification we find that all SNe~Ib/c found 
in early-type galaxies are faint Ca-rich SNe similar to SN 2005E. Note that the SN 2005E-like 
SN host distribution is very different from that of other SNe~Ib/c, 
as well as that of SNe~II (known to have young massive progenitors) 
and that of SN 2002cx-like SNe~Ia, with half of the SN 2005E-like group 
(four out of eight) observed in early-type (elliptical or S0) galaxies. 
The progenitors of SN 2005E and the other members of its group are therefore
likely to belong to an old, low-mass stellar population. The total numbers of 
host galaxies included in this figure are 244, 25, 8, 257, 30, 63, 14, and 8 for 
SNe of types Ia, 91bg, 91T, II, Ib, Ic, 02cx, and 05E, respectively.


\bigskip
\bigskip
\newpage 
\begin{centering}
\underline{\Large Supplementary Information}\\
\end{centering}

~\\

\noindent
{\bf (1) Technical observational details for figures 1 and 2}

\noindent\underline{Figure 1}
Panel (a) shows the image of NGC 1032 prior to explosion of SN 2005E, obtained
from the Sloan Digital Sky Survey (SDSS) archive.
Panel (b) shows the LOSS\cite{fil+01} discovery of SN 2005E on Jan. 13, 2005. 
LOSS imaging of SN 2005E was obtained using the robotic 76-cm Katzman Automatic Imaging Telescope (KAIT) at Lick Observatory.
Panel (c) shows an image of NGC 1032 in the light of the 
H$\alpha$ emission line; the panel shows the difference between images 
obtained using a narrow filter ($6567$~{\AA}; full width at 
half maximum $\sim 100$~{\AA}) with a measured 
transmission of $\sim 40$\% for H$\alpha$ at the redshift of NGC 1032, 
and broad $R$-band observations used for continuum subtraction.
The images, with exposure times of 1800~s, were obtained on Oct. 5, 2008 
using the RATCam camera mounted on the 2-m Liverpool Telescope 
at Observatorio del Roque de Los Muchachos (La Palma, Spain). 
The smooth negative residual ($\sim7\%$ of the original flux) 
near the galaxy core probably arises from a combination of 
slight color gradients
of the smooth galactic old population, and H$\alpha$ absorption in the spectra
of old stars, and does not indicate real line emission.
Panel (d) shows a zoom-in on the location of SN 2005E in pre-explosion SDSS 
$r$-band images. 
Panels (e)-(f) show deep photometry  of the SN 
location obtained using the Low-Resolution Imaging Spectrometer
(LRIS)\cite{oke+95} mounted on the Keck-I 10-m telescope on Feb. 17, 2009
under very good conditions (seeing $\sim0.7''$). 
Panel (e) shows a red image with  a total exposure time of 840~s, 
reaching a point-source detection limit of
$r<25.3(25.9)$ mag at $3(2)\sigma$.
Panel (f) shows a UV ($u$-band) image with a total exposure time of 780~s, reaching a point-source detection limit of
$u<24.7(25.7)$ mag at $3(2)\sigma$. 
Panel (g) shows a zoom-in on the location of SN 2005E in H$\alpha$ light 
from the same observations used to produce panel (c). 

\noindent\underline{Figure 2}
Upper panel: Photospheric spectra of SN 2005E. 
The top two spectra were obtained as part of the Caltech Core-Collapse Project 
(CCCP)\cite{gal+07} using the double-beam spectrograph\cite{oke+82} mounted
 on the 5-m Hale telescope at Palomar Observatory. Exposure times were 600~s 
and 900~s on 2006 January 16 and February 6, respectively, with 
the 158 lines mm$^{-1}$ and 1,200 lines mm$^{-1}$ gratings, yielding an 
instrumental resolution of $\sim5$~{\AA} and $\sim0.5$~{\AA} on the 
red and blue sides, respectively. The CCCP spectra were further rebinned 
to $\sim 5$~{\AA} resolution bins to increase the signal-to-noise
ratio. The bottom spectrum was obtained using LRIS\cite{oke+95}
mounted on the Keck~I 10-m telescope on 2005 March 11.
 We took an exposure of 600~s using the 560 dichroic and the 400/8500 grating  
and 600/4000 grism, giving resolutions of 5.6~{\AA} and 2.4 {\AA} in the red and 
blue sides, respectively.
For comparison, we plot a nebular spectrum of SN 1991bg; see Ref. \cite{maz+97}
for a detailed discussion.  

Note that the nebular spectrum contains residual photospheric 
light. We have addressed the issue of disentangling the 
photospheric and nebular 
components of late-time spectra (in order to use the nebular part for
abundances analysis; Figs. S1, S2) following the approach used in the similar
case of SNe 1997ef and 1997dq\cite{maz+04}. 

Our method of choice was to use the spectral-fitting code {\it Superfit}\cite{how+05}
 to find the best match to the late-time spectrum of SN 2005E with the fit limited to 
a spectral range ($\lambda < 6200$~{\AA}) which is free from strong nebular emission
 lines. The best-fit spectrum was that of the type Ic SN 1990U, 41 days past discovery,
 which provides an excellent fit to the photospheric component (Fig. S1) across the
 observed spectrum including the wavelength ranges not included in the fit. We then 
replaced areas in the best-fit SN 1990U template spectrum affected by residual host-galaxy 
narrow H$\alpha$ line contamination by a linear interpolation, and subtracted
 the resulting photospheric ``best-fit spectrum'' from the SN 2005E data to get a 
clean nebular-only spectrum, which was then used in the nebular-modeling analysis

We note that alternative photospheric subtraction methods (e.g., 
wavelength-unconstrained {\it Superfit} modeling or simple low-order polynomial 
continuum fits) give very similar results. 

\clearpage

\centerline{\includegraphics[width=4.5in,angle=0]{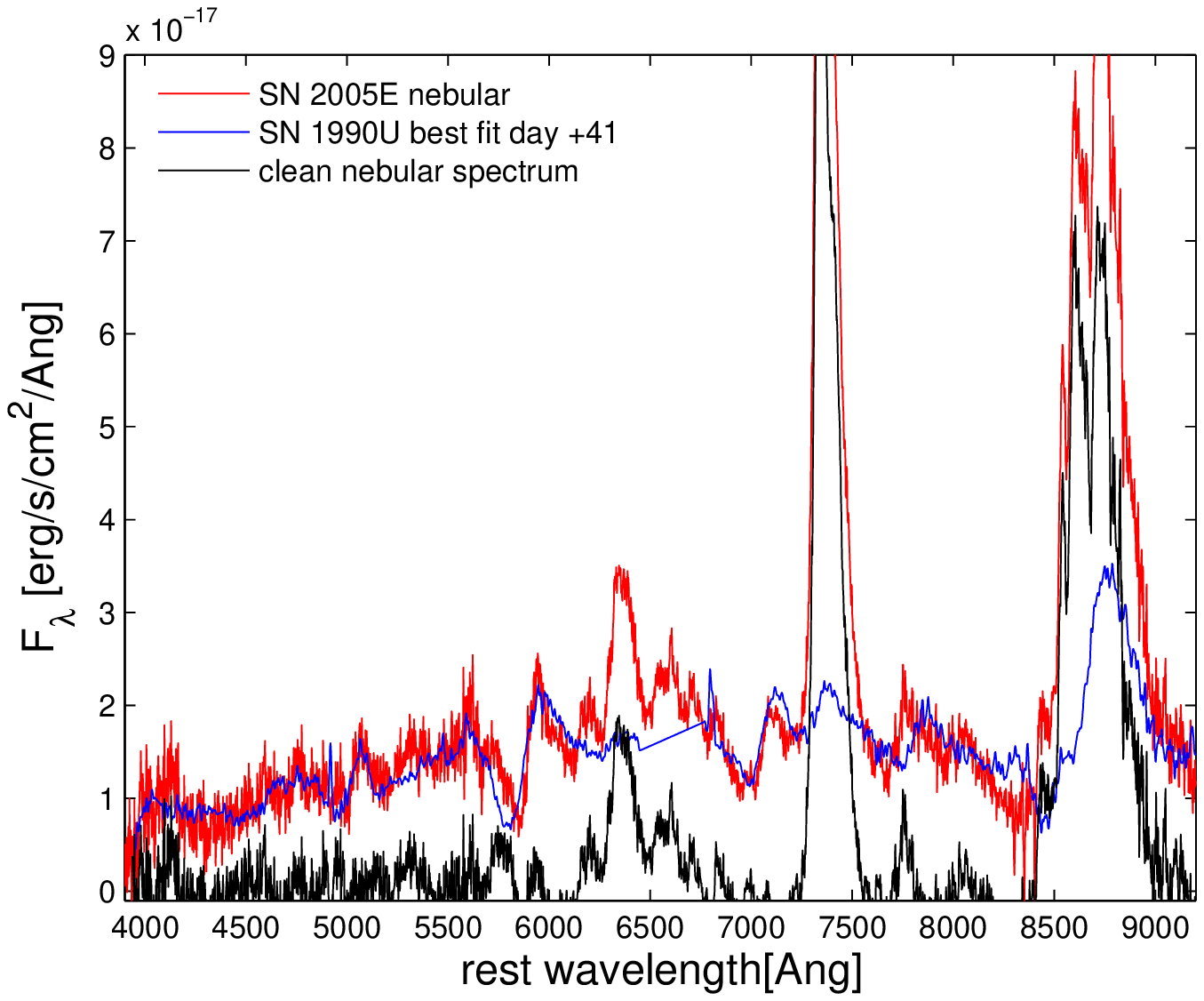}}
\bigskip {}
\noindent{\bf Figure S1}:\\
\begin{scriptsize}
Fitting the photospheric component of the late-time spectrum of SN 2005E. 
We show the best-fit photospheric template (SN 1990U 41 days past discovery, blue)
superposed on the original SN 2005E spectrum (red); note the excellent
agreement outside the range of the strong nebular Ca and O lines. The final 
clean nebular spectrum (black) was obtained by subtracting the best-fit template
(blue) from the data (red) after interpolating over the H$\alpha$ region in the template,
contaminated by narrow host-galaxy lines. 

\end{scriptsize}

\clearpage
\bigskip
\noindent
{\bf (2) Can the progenitor of SN 2005E be a massive star formed in a star-forming region in the halo?}

Massive stars are typically formed and observed in 
giant molecular clouds and young stellar clusters or 
associations\cite{chu+08,sch+08}. Core-collapse SNe from massive stars are 
therefore expected to be found close to star-forming regions. 
Observations of such SNe are usually  
consistent with this picture\cite{and+08}.
In principle, the discovery of SN 2005E in the halo of NGC 1032 could
be attributed to in-situ star formation of a massive star rather than 
a low-mass older progenitor. However, star formation
in the halo environment of a S0/a galaxy would be difficult to understand
according to current star-formation theories. For example, star formation 
during collisions between cloudlets within
high-velocity clouds at high galactic latitudes\cite{dys+83} has been 
shown to be much too rare\cite{chr+97}. Spiral density waves 
in the disk may trigger star formation  up to a kpc
above the Galactic plane\cite{mar+99}, but this seems unlikely 
for the larger height of SN 2005E (which also appears
to be positioned beyond the edge of the optical disk). In addition, 
we note that NGC 1032 shows no evidence
for warping or other structures extending beyond the region of the
galactic disk to which SN 2005E could be related.
We conclude
that given the remote location of the supernova in the galactic
halo, and the nondetection of any star-formation activity anywhere
in the halo or the disk of NGC 1032, it is unlikely that an in-situ
formation scenario could explain SN 2005E, unless a yet unknown and
unique star-formation mechanism was at work in this case. 
In contrast, the evidence for a low-mass progenitor of SN 2005E is naturally 
consistent with the low-mass old stellar population environment 
in which it was found. Such evidence is further supported by the host-galaxy distribution 
of the larger sample of Ca-rich SNe we report here, which is dominated by early-type galaxies 
(see Fig. 3).  

\noindent\underline{A search for nearby star-forming regions:}

We have looked for star-formation tracers both in the halo and the
disk of NGC 1032. Star-forming regions 
produce two classes of emission: continuum emission from young
stars and emission lines (dominated by H$\alpha$) produced by ionized gas. 
We have searched for both classes of emission, and obtained upper
limits on the star-formation rates.  

\noindent\underline{H$\alpha$ observations:}

H$\alpha$ imaging was obtained with the Liverpool 
Telescope, and then analyzed using similar methods to those described in detail 
elsewhere\cite{and+08}. We have determined an upper limit 
of $2.02\times10^{-17}$ erg cm$^{-2}$ on the H$\alpha$ flux from the region 
of SN 2005E. This is a 3$\sigma$ upper limit obtained from the variation in the 
sky background, for a $2''$  aperture centred on the SN position and 
calibrated using an $R$-band galaxy magnitude taken from the literature\cite{her+96}. 
For the distance of NGC 1032 (34 Mpc) we infer an H$\alpha$ luminosity of
$2.79\times10^{36}$ erg s$^{-1}$; correcting this for Galactic extinction 
(0.098 mag) and for the contribution from [N~II] lines\cite{ken+83}, 
we then calculate a corrected limit of 
H$\alpha_{\rm limit} = 2.3\times10^{36}$ erg s$^{-1}$. Using the conversion rate 
from Ref. \cite{ken98} (Eq. 2), we determine an upper limit on the 
star-formation rate (SFR) at the SN position of ${\rm SFR}_{\rm limit}= 1.8\times10^{-5}$ 
M$_{\odot}$ yr$^{-1}$ down to our detection limits. 

In addition, our H$\alpha$ observations of NGC 1032 show no star-forming regions closer
than the galactic nucleus itself (see Fig. 1; likely the origin of the H~I detected 
in this galaxy\cite{spr+05}, similar to many other such galaxies in which molecular 
gas is typically centrally concentrated\cite{wel+03}), up to our detection limit.

\noindent\underline{$R$- and $u'$-band observations:}

Our deep $R$ and $u'$-band observations using Keck (calibrated onto the SDSS
photometric system) rule out point sources near the location of SN 2005E down to 
$u'<24.7\,(25.7))$, $r<25.3\,(25.9)$ mag at 3(2)$\sigma$ (see Fig. 1). At the distance of 
NGC 1032, these limits 
($M_r<-7.5(-6.9)$ and $M_{u'}<-8.1(-7.1)$ mag at $3(2)\sigma$, respectively) indicate
that we would have detected faint star-forming galaxies or star-forming regions 
at the SN location, or indeed even individual red supergiant or luminous 
blue supergiant stars (although the existence of a single $<20$~M$_\odot$ 
star cannot be ruled out). 
Since massive stars are usually formed and observed in 
young stellar clusters or associations\cite{chu+08,sch+08}, the lack of nearby supergiants 
(either red or blue) further argues against local star-formation activity. 
The progenitor of SN 2005E could have been a member of an undetected globular cluster; 
the stellar population in such a hypothetical cluster, however, would be very old (several Gy), 
again indicating an old, low-mass progenitor for SN 2005E.

\noindent\underline{Radio signature of a core-collapse SN:}

A non-negligible fraction of core-collapse SNe show radio emission. 
We have therefore made observations at 8.46 GHz with the VLA radio telescope 
on Jan. 21.10, 2005. We found
a flux of $11\pm53\,\mu$Jy at the optical position of SN 2005E.
At the distance of SN 2005E, and assuming an explosion date between 
Dec. 24, 2004 and Jan. 14, 2005, the radio luminosity limit ($2\sigma$) is $1.8\times10^{26}$ erg s$^{-1}$ Hz$^{-1}$,
 which is a factor of 10 lower than a typical radio-emitting SN Ib/c on this same
timescale\cite{sod07}.

\bigskip
\noindent{\bf(3) Spectroscopic identification of SN 2005E as a type Ib supernova}

In Fig. 2 we show optical spectra of SN 2005E. Our first 
spectrum (Fig. 2 top, blue curve) is clearly photospheric,
dominated by absorption lines including the He~I series
at 4471, 5876, 6678, and 7065 {\AA}, blueshifted
by $\sim11,000$ km s$^{-1}$ (marked with black ticks
at the top of Fig. 2), typical of a young SN~Ib.
Based on prediscovery nondetections, SN 2005E was 3--20 days
after explosion at this time. Analysis using the {\it Superfit} 
spectral analysis code\cite{how+05} confirms a 
type Ib identification, with the best-fit match being with a 
spectrum of the type Ib/c transition event SN 1999ex\cite{ham+02}
14 days after maximum light.

Our next spectrum (Fig. 2 middle, red curve) 
shows the beginning of
the transition to the nebular phase, with emerging emission lines
of calcium. The best-fit spectrum found by {\it superfit}
is that of the type Ic event SN 1990U, but spectra of the type Ib SN 1999di
also provide a good fit. Masking of the blueshifted He~I 6678 {\AA} line by
the emerging [O~I] $\lambda\lambda$6300, 6364 nebular doublet  
may account for the similarity to SNe~Ic, with intrinsically much 
weaker He~I lines. 

Strong nebular emission lines of [O~I] $\lambda\lambda$6300, 6364,
and especially [Ca~II] $\lambda\lambda$7291, 7324 and the 
Ca near-infrared triplet at 
8498, 8542, and 8662~{\AA} (tick marks at the bottom of Fig. 2), 
dominate our latest spectrum of SN 2005E (Fig. 2  bottom, magenta curve). 
The best fit found by {\it Superfit}
is to the type Ib SN 1985F\cite{fil+86,gas+86} 
obtained 89 days after its first observation. 
Dominated by lines of intermediate-mass elements (O and Ca), 
the nebular spectrum of SN 2005E is similar to that of SNe~Ib,
though calcium is much stronger than usual for this SN type,
while the lack of Fe-group emission lines in the bluer part of the spectrum
rules out an identification as a SN~Ia of either the normal or subluminous 
(SN 1991bg-like\cite{fil+92}) varieties. 

\bigskip
\noindent{\bf(4) The relative ejected mass of calcium and sulfur}

To constrain the relative ratio of ejected calcium to sulfur (Ca/S, used 
below to investigate the nucleosynthetic processes responsible for 
SN 2005E-like events), we ran a grid of nebular spectral models 
similar to those shown in Fig. 2b. However, we forced the S/Ca 
fractions to be 0.16, 0.37, 0.55, 0.74, and 1.00 as shown in Fig. S2. 
As can be seen there, the data rule out ratios below Ca/S = 6 
(this limit is set by the uncertainty in photospheric light subtraction
 and blending with weak nebular C lines). Note that the similar excitation 
parameters and critical densities of [S~I] $\lambda\lambda$4589, 7725 
and [O~I] $\lambda\lambda$6300, 6364 (which is well detected) 
indicate that the lack of strong [S~I] lines results from an abundance 
(rather than an excitation) effect. 

\clearpage

\centerline{\includegraphics[width=3.5in,angle=0]{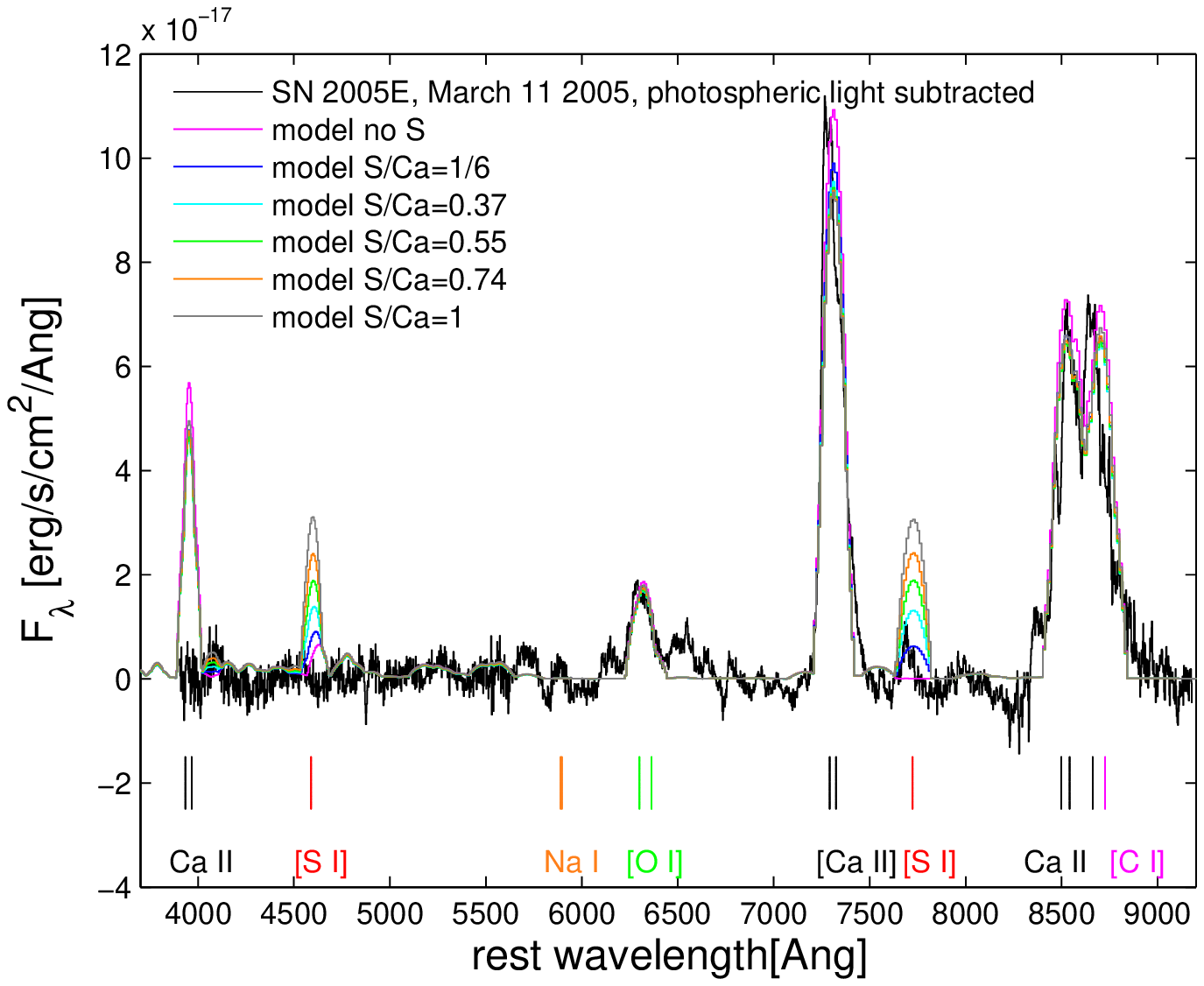}}
\centerline{\includegraphics[width=3.5in,angle=0]{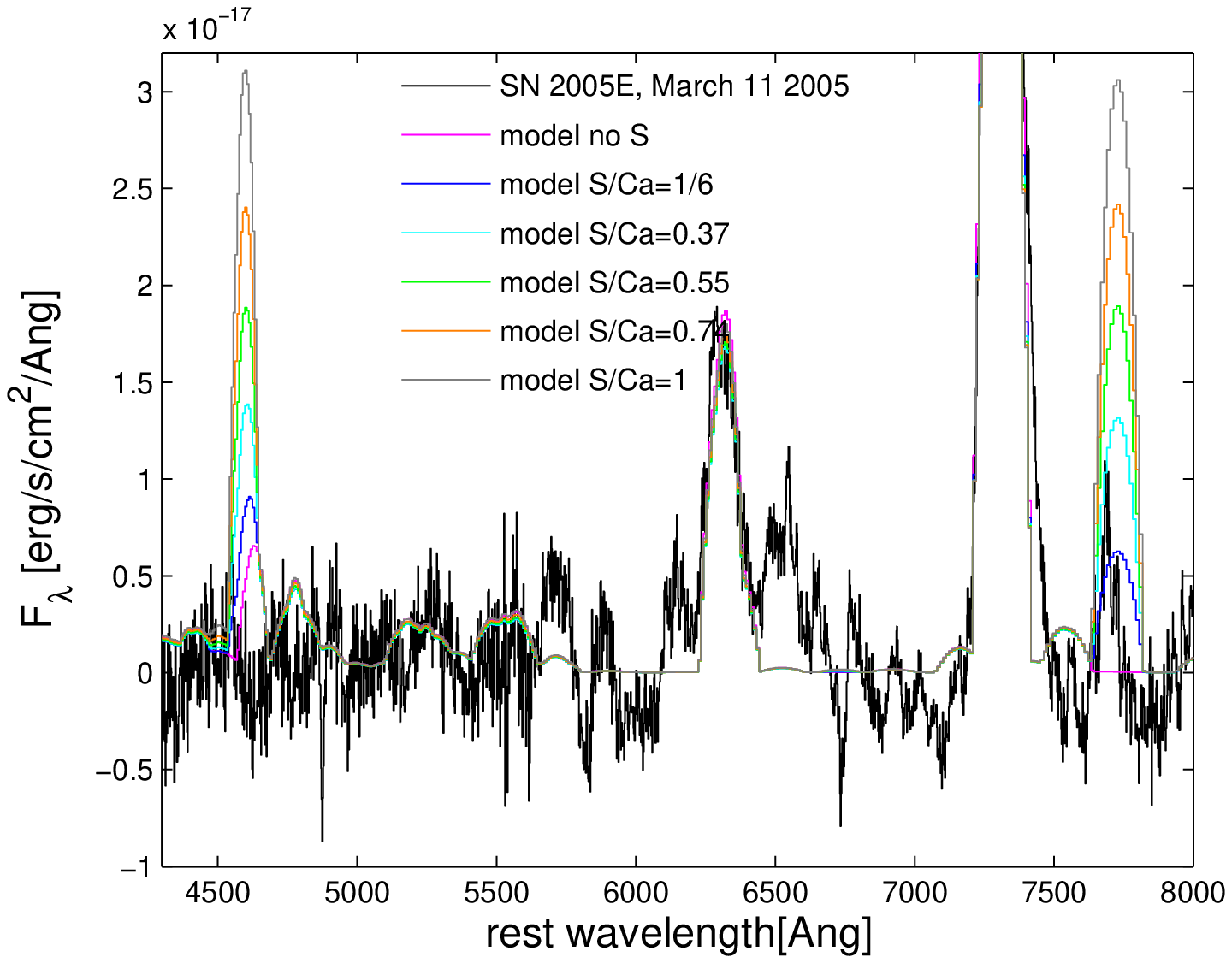}}
\bigskip 
\noindent{\bf Figure S2}:\\
\begin{scriptsize}
An upper limit on the relative abundance of sulfur and calcium. 
Top: a grid of models calculated using the same input spectrum,
 with values of the S/Ca ratio artificially set to 1/6, 0.37, 0.55,
 0.74, and 1.00. Note that values above S/Ca = 1/6 are clearly ruled out
 by the data. S lines are not detected, but lower values cannot be
 constrained due to uncertainty in the photospheric light 
subtraction and blending with nearby C lines. The bottom panel shows 
that the [O~I] $\lambda\lambda$6300, 6364 line, whose excitation parameters
and critical densities are similar to those of the [S~I] $\lambda\lambda$4589, 
7725 lines, is well detected, and thus the lack of S lines is a real abundance 
effect rather than an excitation effect.  
\end{scriptsize}
\clearpage 

\bigskip
\noindent{\bf(5) Masses and luminosities of SNe}

The total mass, nickel mass, and luminosity of SN 2005E are far 
lower than those found for the majority of SNe of any type (Fig. S3). 
Taken together, no other SN except for SN 2008ha (see SI, Section 10 
for a discussion of this SN and other related events) was both faint 
and has subsolar-mass ejecta. SNe~Ia (excepting SN 2008ha), both 
regular and peculiar/subluminous, have inferred ejecta masses of 
$\ge \sim1$~M$_\odot$\cite{maz+07}. SNe~Ib/c and SNe~II have even 
more massive ejecta (a few times 
solar)\cite{nad03,maz+00,tom+05,zam05,li06,sau+06,maz+07b,utr+07}. 
Additional data concerning the ejected Ni mass can be found in 
Ref.\cite{sod+06a}. Some SNe~II-P are as faint 
as SN 2005E or even fainter\cite{pas+04}, but ejected a few solar masses. 

\bigskip

\clearpage

\centerline{\includegraphics[width=5.5in,angle=0]{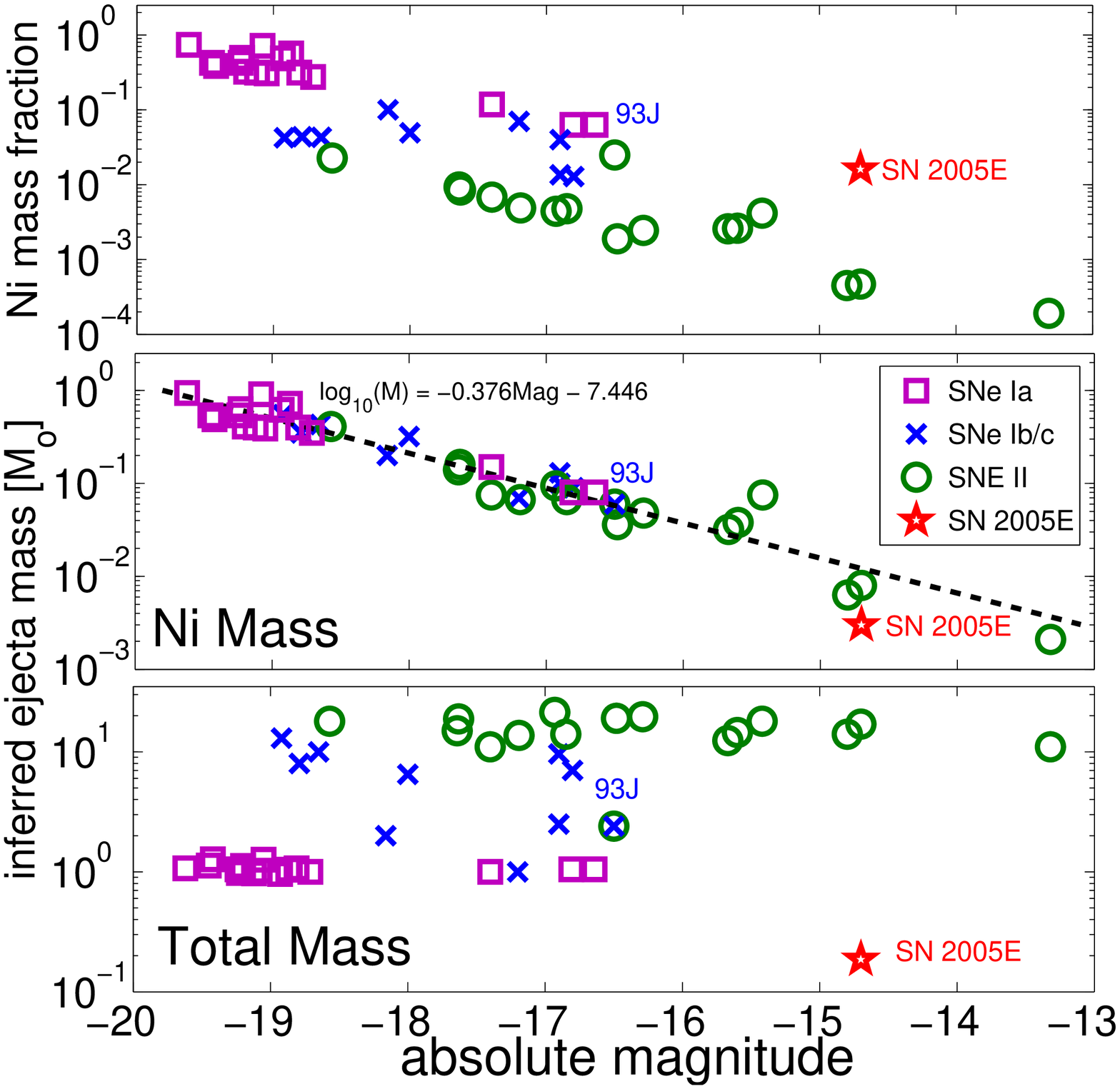}}
{\bf Figure S3}:
\begin{scriptsize}
Comparison of the SN 2005E ejecta mass and luminosity with those of 
other SNe [SNe~Ia, squares; SNe~Ib/c, $\times$ marks; SNe~II, circles]. 
The lower panel shows the total ejecta mass inferred for SN 2005E, 
which is the lowest inferred ejecta mass found for any SN, based on 
nebular spectra. Its position
in the luminosity vs. ejecta-mass phase space is unique, suggesting 
that it is not a member of the currently well-known SN families. The 
middle panel shows the Ni mass inferred for SN 2005E; it is small,
consistent with its low luminosity, although somewhat lower than 
might be expected from the extension of the observed Ni mass vs. 
luminosity relation observed for other SNe (dashed line and formula). 
The upper panel shows the Ni mass fraction, $M_{\rm Ni}/M_{\rm total}$, 
inferred for SN 2005E. 
All of the masses in this figure were inferred from detailed 
modeling. The error bars on the masses are $\sim 10$--20\%; these are 
smaller than the symbol sizes and are not shown.
\end{scriptsize}

\noindent{\bf(6) Ejected mass estimates from the observed light curve and photospheric velocities}

The ejecta mass of a given supernova can be estimated using its light curve and the
observed ejecta velocities. 
The expansion velocity, $v$, of a SN is proportional to $(E_{\rm kin} 
/M_{\rm ej})^{1/2}$, where $E_{\rm kin}$ is the kinetic energy and 
$M_{\rm ej}$ is the ejected mass, while the typical duration of a SN light 
curve is $t_{\rm d}\propto (M_{\rm ej}^{3} /E_{\rm kin})^{1/4}$ \cite{arn82}.  
Combining these equations and
assuming that two objects have the same opacity, we have
\begin{equation}
  E_{{\rm kin, } 1} / E_{{\rm kin, } 2} = \left ( \frac{v_{1}}{v_{2}}
    \right )^{3} \left ( \frac{t_{1}}{t_{2}} \right )^{2}
\end{equation}
and
\begin{equation}
  M_{{\rm ej}, 1} / M_{{\rm ej, } 2} = \frac{v_{1}}{v_{2}} \left (
    \frac{t_{1}}{t_{2}} \right )^{2}.
\end{equation}
\noindent

Following Ref. \cite{fol+09}, in which the mass of the subluminous 
SN 2008ha was estimated, we use
a normal SN~Ia as a reference  
with $t_{\rm d} = 19.5$~days and $v =8,000$~km~s$^{-1}$\cite{ste+05}. 
The timescales of SN 2005E are a factor of 0.45--0.55 times 
those of the well-observed type Ib SN 2008D (the contracted light 
curve of SN 2008D is shown in Fig. S4 for comparison), which
had a rise time of 18 days\cite{mod+09}, and therefore we estimate 
the rise time of SN 2005E to be 7--9 days. The ejecta velocities we 
observe from the photospheric spectra are 11,000 km s$^{-1}$. 
We therefore find $E_{\rm kin, 05e} / E_{\rm kin, Ia} = 0.34$--0.55 
and $M_{\rm ej, 05e} / M_{\rm ej, Ia} = 0.17$--0.29.
Assuming $E_{\rm kin, Ia} = 1.3\times10^{51}$~ergs and $M_{\rm ej, Ia} = 1.4$
M$_{\odot}$, we find $E_{\rm kin, 05e} = (4.4-7.2) \times 10^{50}$~ergs and
$M_{\rm ej, 05e} = 0.25-0.41$ M$_{\odot}$. We note that this widely used 
method for SN ejecta-mass estimation may have two caveats in our case:
helium is less opaque than other elements so we may miss
some of the helium mass, and $^{56}$Ni may not be the only radioactive
energy source as assumed in such estimates.   

\clearpage

\centerline{\includegraphics[width=3.5in,angle=0]{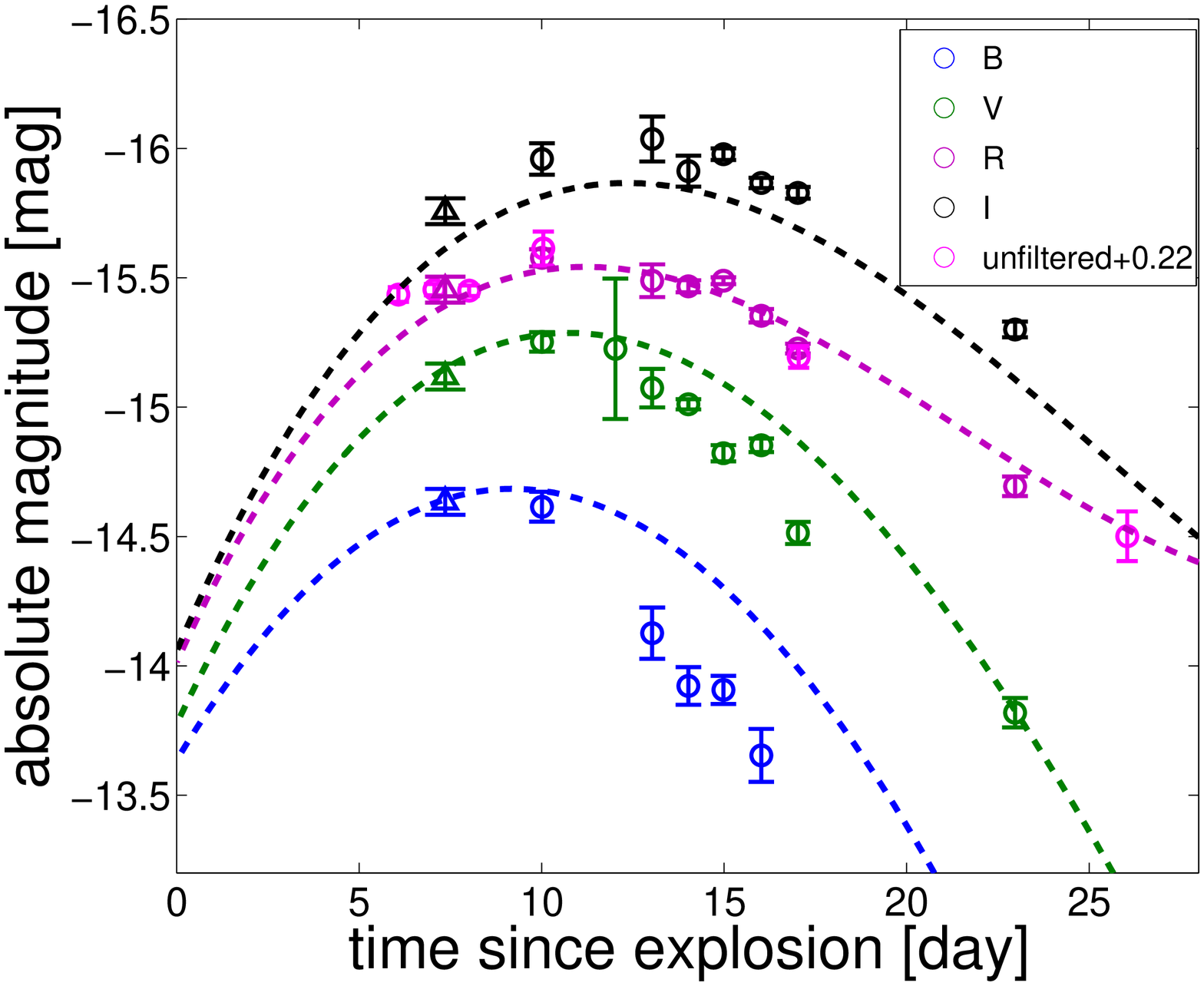}}
\centerline{\includegraphics[width=3.9in,angle=0]{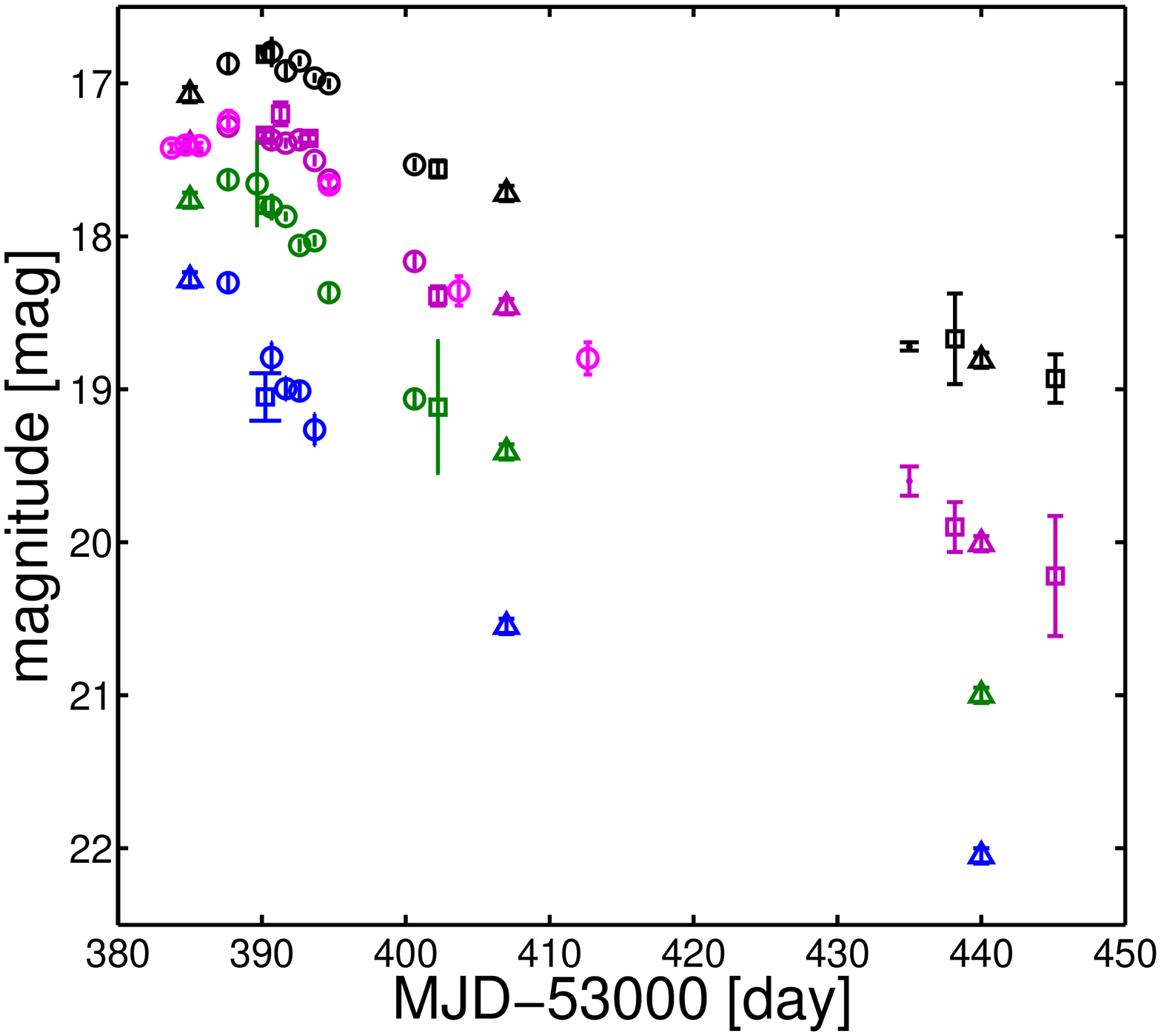}}
\bigskip 
\noindent{\bf Figure S4}:\\
\begin{scriptsize}
Optical light curves of SN 2005E. We present our observations obtained using
the 0.76-m KAIT as part of LOSS\cite{fil+01} in the $BVRI$ bands 
(blue, green, red, and black empty circles, respectively),
as well as unfiltered data (magenta empty circles). Also shown
are CCCP\cite{gal+07} $BVRI$ observations obtained using the Palomar 
60-inch telescope [P60;\cite{cen+06}; (rectangles)] and $RI$ photometric 
observations with Keck
(red and black points, respectively). Additional points (triangles)
 were obtained through synthesis of the spectra 
(seen in Fig. 2), scaled to fit the $R$-band photometry.  
KAIT unfiltered observations are most similar 
to $R$-band data due to the combined response of its optics and detector,
so we have scaled these data to the $R$-band observations. 
Top: Light curve near peak. The rapid decline of this object is consistent
with its relatively low
absolute magnitude, $M_R \approx -15.5$, calculated assuming a distance of 
34 Mpc to NGC 1032 (as given in the NED database\cite{ned}) 
and negligible extinction. Estimating the peak date
of this SN from the magenta curve (January 7, 2005), our spectra (Fig. 2) were obtained 
9, 29, and 62 days after maximum light. 
Also shown for comparison are third-order polynomial fits to the light curves of the 
well-studied type Ib SN 2008D\cite{mod+09} time-contracted by a factor
of 0.52 to fit the SN 2005E data.
 We find that SN 2005E behaves similarly 
to SN 2008D near peak, but with timescales multiplied by roughly
0.45--0.55.  
Bottom: long-term evolution of the light curve. The rapid light-curve decline 
is similar to that observed in subluminous SNe~Ia at early times, 
with a slope of $-0.155$ mag day$^{-1}$; however,
it does not show a break to a 
different slope expected at $\sim4.5$ days after peak  
(compare with the analysis of Kasliwal et al.\cite{kas+08}),
but continues instead to decline at a nearly constant rate, even at
late times.    
\end{scriptsize}
\clearpage

\noindent{\bf(7) Composition of SN ejecta}

Table 1 shows the elemental abundances inferred for various SNe.
SN 2005E synthesized more Ca than any other known SN in the absolute sense,
by typically a factor of 5--10, with the exception of massive hypernovae 
like SN 1998bw. Even these rare extreme events lag behind in the Ca mass 
fraction (a factor of 30 less than in SN 2005E) while typical SNe have 
Ca mass fractions 
of $>100$ times below that of SN 2005E, so this class indeed merits 
the ``Ca-rich'' distinction initially proposed by Ref.\cite{fil+03}.
\begin{table}
\begin{scriptsize}
\begin{tabular}{|c|c|c|c|c|c|c|c|c|}
\hline 
SN (Refs.)/Element & C & O & Mg & Si & S & Ca & Ni\tabularnewline
\hline
\hline 
1998bw\cite{tan+07} & 0.15 (1.9) & 3.9 (49) & 0.05 (0.63) & 0.15 (1.9) & 0.4 (5) & 0.08 (0.015)& 0.65 (8.2)\tabularnewline
\hline 
1997dq\cite{maz+04} & 0.37 (9.2) & 0.73 (18) & 0.003 (0.075) & 0.8 (20) & 0.2 (5) & 0.04 (0.017) & 0.16 (4)\tabularnewline
\hline 
2002ap\cite{maz+07b} & 0.4 (40) & 1.03 (103) & 0.01 (1) & 0.25 (25) & 0.1 (10) & 0.01 (0.005) & 0.11 (1.1)\tabularnewline
\hline 
2006aj\cite{maz+07c} & 0.25 (1250) & 1.42 (7100) & 0.003 (15) & 0.05 (250) & 0.01 (50) & 0.0002 (0.0001) & 0.19 (950)\tabularnewline
\hline 
1994I\cite{sau+06} & 0.11 (28) & 0.33 (83) & 0.003 (0.75) & 0.1 (25) & 0.05 (12) & 0.004 (0.006) & 0.08 (20)\tabularnewline
\hline 
2003du \cite{maz+07c}& 0 (0) & 0.23 (230) & 0.022 (22) & 0.21 (210)  & 0.08 (80) & 0.001 (0.0009) & 0.59 (590)\tabularnewline
\hline 
2004eo\cite{sau+08} & 0 (0) & 0.43 (86) & 0.038 (7.6) & 0.26 (52)  & 0.04 (8) & 0.005 (0.0045) & 0.32 (64)\tabularnewline
\hline 
1991bg\cite{maz+97} & 0 (0)  & 0.24 (120) & 0.04 (2) & 0.87 (435) & 0.02 (10) & 0.002 (0.0016) & 0.08 (40)\tabularnewline
\hline 
2005hk\cite{sah+08} & $<$0.014 (2.5) & 0.53 (132) & -- & 0.02 (5) & 0.022 (5.4) & 0.004 (0.004) & 0.4 (100)\tabularnewline
\hline 
2005E & 0.1 (0.74) & 0.037 (0.27) & 0 (0) & 0 (0) & $<$0.01 (0.074) & 0.135 (0.49) & 0.003 (0.022)\tabularnewline
\hline
\end{tabular}

\caption{\label{t:abundances} \footnotesize Elemental abundances in
  various SNe (abundance ratios relative to Ca are shown in brackets;
  for Ca the ratio shown is to the total mass inferred for the SNe
  ejecta). This sample includes SNe of types Ib, Ic, and Ia. The
  SNe~Ia include the prototypical subluminous SN 1991bg as well as SN
  2005hk (the latter is a member of the SN 2002cx-like group). All
  abundances were derived using the same methods and in most cases
  using the same code, such that any possible systematic problems
  cancel out. Note that we estimate a conservative error of $<25$\% on
  the calcium abundance in SN 2005E. Specific details are given in the
  references for each of the SNe. }
\end{scriptsize}
\end{table}
\newpage

\bigskip
\noindent{\bf(8) The possible origin of SN 2005E from a hypervelocity star and its trajectory}

 Most of the stars in the Galaxy move at relatively low velocities
(a few tens of km s$^{-1}$) with respect to their Galactic environment.
Some massive stars ($\simgt 8$ M$_\odot$) are known
to have higher velocities, up to $\sim200$ km s$^{-1}$. These so-called 
``runaway stars" are ejected from their birth place as a result
of binary encounters, or if they have binary companions that explode
as SNe\cite{bla61,hoo+01}, and could therefore be found
far from the star-forming region where they were formed. Hypervelocity
stars\cite{hil88} (HVSs) move at even higher velocities (few$\times10^{2}$--$10^{3}$
km s$^{-1}$). Such stars are thought to be ejected following dynamical
interaction with (or close to) massive black holes.

In 2005, the first HVS was serendipitously discovered in the Galactic halo\cite{bro+05},
 $71$ kpc from the Galactic centre (GC),
with a radial velocity of $853\pm12$ km s$^{-1}$. Two additional
HVSs, one of them a massive B star ($\sim9$ M$_\odot$),
were discovered shortly thereafter\cite{ede+06,hir+05}. Follow-up surveys
have discovered a total of $\sim20$ HVSs with radial velocities in the range
300--900 km s$^{-1}$ at distances of 20--120 kpc from the GC. 
A total population of $\sim100$ such young B stars is inferred
to exist in the Galaxy at these distances (no main-sequence O stars are observed). 
Given their positions and velocities, all observed HVSs must have 
been ejected with even higher initial velocities ($>850$ km s$^{-1}$), if they originated near
 the massive black hole (MBH) in the centre of our Galaxy whose mass is 
 $\sim3.6\times10^{6}$ M$_\odot$\cite{eis+05,ghe+05}.

 If the progenitor of SN 2005E was a massive star, it may have been formed in the centre 
or in the disk of NGC 1032, and was later ejected at high velocity, traveling to its observed location 
in the halo.
 In order to find the ejection velocity required for the progenitor of SN
2005E to travel from its birth place to the observed position
of SN 2005E, we must trace its possible trajectory. For this purpose
 we need to assume some galactic potential for NGC 1032 as well as the mass for the
MBH in its nucleus.

  For the potential of NGC 1032, we use a two-component model suggested in Ref. 
\cite{miy+75} composed of a galactic bulge and a halo (the disk component has a relatively
 small effect). The bulge mass could be estimated from the velocity dispersion in the
bulge\cite{gor+07} to be between $6\times10^{10}\, \rm{M}_{\odot}$ 
and $1.3\times10^{11}\, \rm{M}_{\odot}$, where the bulge size is $\sim1.85$
kpc\cite{gor+07}. Given the velocity dispersion in the bulge of NGC 1032, which is in
the range\cite{gor+07} 200--225 km s$^{-1}\,$, we can also estimate
the mass of the MBH in the nucleus using the $M-\sigma$ relation\cite{fer+00,geb+00} to be 
in the range $\sim$(1--2) $\times10^{8}\, \rm{M}_{\odot}$. 
 For the halo mass we use Ref. \cite{pra+03}, which finds the virial mass of galaxies such
 as NGC 1032 to be $\sim1.5\times10^{12}\, {\rm M}_{\odot}$ for magnitudes
$-19.5 < M_{B} < -20.5$ (NGC 1032 has $M_{B} = -19.8$ mag).  This mass
 is also generally consistent with the relation between halo
mass and velocity dispersion obtained in Ref.  \cite{sha+06} (see their Figure 3c), 
which gives a total galactic mass of (2--3) $\times 10^{12}\, \rm{M}_{\odot}$. 

The inferred ejection velocity of the progenitor of SN 2005E from a birth
place in the disk (taking the shortest distance from the observed
position of SN 2005E to the plane of the galactic disk) is found to
be $>1600\, \rm{km}\, \rm{s}^{-1}$ for a massive ($\sim25\,\rm{M}_{\odot}$), 
short-lived ($\sim7\times10^{6}$ yr) progenitor, appropriate for a SN~Ib. 
Such an ejection velocity
from the galactic disk would require an ejection mechanism different
from that of OB runaways, where the only suggested mechanisms involve
an interaction with a MBH. Such a MBH is unlikely to exist in the
galactic disk. An ejection velocity of $>300$ km s$^{-1}$ would
be required for a lower mass ($\sim8\, \rm{M}_\odot$) and longer lived
($\sim4\times10^{7}$ yr) component in a binary progenitor ejected
from the disk. Such a velocity is much higher than the typical velocities
of OB runaway stars\cite{hoo+01}, but it could theoretically be
accessible for a runaway single star\cite{leo+91}. However, runaway
binaries are typically ejected at lower velocities (up to $0.3-0.4$
of the maximal ejection velocities of single stars\cite{leo+90}),
which would again suggest a different ejection mechanism than OB runaways.
We conclude that the progenitor of SN 2005E is unlikely to be ejected from
the disk by currently suggested high-velocity ejection mechanisms,
and (if ejected at all)  was more likely to be ejected from the centre of NGC 1032.
 
For the progenitor of SN 2005E to be ejected from the galactic centre
and reach its current position in the halo during its lifetime, the
required ejection velocity would be at least $\sim3400$ km s$^{-1}$
($\sim1600$ km s$^{-1}$) assuming a lifetime of $<7$ Myr for a $25\, \rm{M}_{\odot}$
star ($<40$ Myr for an $8\, \rm{M}_{\odot}$ binary star). 
We now calculate whether such ejection velocities 
are likely, taking into account the conditions in NGC 1032.

We compare the velocities derived from the trajectories with
the average ejection velocity of a star ejected from the galactic nucleus 
following the disruption of
a binary by a MBH, given by\cite{hil91,bro+06c}
\[v_{\rm eject}=3400\, \rm{km}\, \rm{s}^{-1}\times\left(\frac{a_{bin}}{0.8\, 
\rm{AU}}\right)^{-1/2}\left(\frac{M_{\rm bin}}{50\, 
\rm{M}_{\odot}}\right)^{1/3}\left(\frac{M_{\rm BH}}{1.5\times10^{8}\, 
\rm{M}_{\odot}}\right)^{1/6},\]
where $a_{\rm bin}$ is the semimajor axis of the binary, $M_{\rm bin}$ 
is the binary mass, and $M_{\rm BH}$ is the MBH mass. Massive binaries
usually have components of comparable and frequently equal mass, and
are known to have relatively compact orbits, with a large fraction
of them $(f_{\rm cbin} \approx 0.4$) in close binaries 
($a_{\rm bin}<1\, \rm{AU}$; e.g., Refs. 
\cite{abt83},\cite{mor+91},\cite{kob+07}). 
We therefore conclude that the observed
position of SN 2005E is consistent with its progenitor being ejected
 as a massive star following the disruption of a typical very massive
 binary of $M_{\rm bin}=2\times25=50\, \rm{M}_{\odot}$ and 
$a_{\rm bin}\simlt0.8\, \rm{AU}$,
 or the disruption of a $M_{\rm triple}=3\times8=24\, \rm{M}_{\odot}$ triple
 star, with an outer semimajor axis of $a_{\rm bin}\simlt1.5\,$AU, which
 would eject a hypervelocity binary\cite{per09b}. 

 The velocity of an observed supernova is difficult to measure. The
 measured velocities of the supernova ejecta are on the order of a
 few thousand km\, s$^{-1}$ and the pre-explosion
 velocity of the progenitor is negligible, even for velocities as high
 as hundreds of km\, s$^{-1}$. Even for a HVS, 
 one would require extreme velocities, directed along the line of sight, in order to identify 
 a significant signature of the motion of the SN progenitor. The velocities we find for the
 ejecta of SN 2005E are not unusual, and are consistent with other SNe~Ib/c. 

\bigskip

\noindent{\bf(9) The ejection rate of hypervelocity stars}

Although two massive HVS ($>8\, \rm{M}_{\odot}$; Refs. 
\cite{ede+06},\cite{heb+08})
are known in our Galaxy, it is difficult to infer the total number
of Galactic massive HVSs from the very few examples known, given their
serendipitous discovery nature. An estimate can be obtained if most massive
HVSs in our Galaxy have been ejected through the Hills binary disruption
mechanism\cite{hil88,yuq+03} (which is the likely case\cite{per+07,per09,per+09}). 
In this scenario,
the binary companions of ejected HVSs should have been captured into
close orbits around the MBH\cite{gou+03,yuq+03}; 
the number of such stars should therefore
reflect the number of similar HVSs in the Galaxy. Currently, a few
tens of main-sequence B stars are observed in such close orbits ($<0.04$
pc from the MBH; e.g., Ref. \cite{gil+09}). Approximately half of these stars
(with identified stellar types) are found to be B0--2~V main-sequence
stars, most likely with masses $>8\, \rm{M}_{\odot}$. Given the trend
of massive binaries to have equal-mass components\cite{abt83}, one
can then infer a total of $\sim$10--20 such massive HVSs in our Galaxy.
The total number of $>8\, \rm{M}_{\odot}$ stars in the Galaxy is $\sim10^{6}$
(e.g., assuming a Miller-Scalo initial mass function); thus, the
HVS fraction of the population of massive stars is $\sim10/10^{6}=10^{-5}$, 
and we therefore expect a similar fraction of HVS supernova
 progenitors.  The rate of hypervelocity binary ejection is likely to be 
a factor of 20--100 times lower than that of single HVSs\cite{per09b}, making 
this possibility highly improbable.  
If the progenitor of SN 2005E were a hypervelocity single star, 
the probability of discovering SN 2005E in LOSS, 
which detected a total of $\sim550$ core-collapse SNe from 1998 through 2008,
would be low. Its discovery would then be either a chance observation of a 
rare event,
or suggests a much higher ejection rate of extragalactic HVSs than observed in
our Galaxy. However, given the low-mass ejecta observed for SN 2005E, its 
additional peculiarities (which cannot be explained by a 
hypervelocity progenitor), and the seven additional SNe found to have similar
properties (three of them in elliptical galaxies), it is unlikely that 
such explosions had hypervelocity massive stellar progenitors.  

\newpage   

 \noindent{\bf(10) Comparison to SN 2002cx-like SNe}

The low luminosity and calcium-rich late-time spectra 
of the recently discovered SN 2008ha (see Fig. 7 of Ref. \cite{fol+09})
have some resemblance to the group of faint calcium-rich SNe presented here
 and could therefore suggest a common origin. However, 
SN 2008ha shares most of its observed properties  
with a group of peculiar SNe~Ia \cite{fol+09,val+09} 
(referred to as SN 2002cx-like SNe after the first 
example observed\cite{li+03}), which are clearly distinct from the new type of 
explosion presented here, in almost all its observed characteristics\cite{jha+06} 
and derived properties (e.g., mass and abundances; 
see Table 2 for a summary of observed differences).
In particular, none of the defining criteria of the SN2002cx-like group
(as described by Ref. \cite{fol+09} and references therein) are observed 
for any member of the SN 2005E-like group. Specifically, none of them 
were observed to have a SN~Ia spectral classification,
very low ejecta velocities (as reflected in the line widths), 
early spectral resemblance to SN 1991T,
slow light-curve evolution, or even the typical luminosity of 
2002cx-like SNe which is $\sim 2$ mag brighter (typical absolute 
magnitude $-17$) than that of SN 2005E and the other members of its 
group ($-15$ mag). Sahu et al.\cite{sah+08} find that
SN 2005hk, the best-observed member of the SN 2002cx-like family, requires 
a mass of C/O products (including Si, S, and Fe) which is of order the 
Chandrasekhar mass. In contrast, the observation of large fractional calcium 
abundances (see Table 1), 
especially without the
accompanying Si, S, and Fe (see SI, Section 4), distinguishes SN 2005E, 
and requires 
a different physical explosion mechanism --- one involving helium burning.
 The physical distinction between the two groups is bolstered by 
the different stellar environments in which these events occur 
(see Fig. 3).

\clearpage
\begin{table}
\begin{scriptsize}
\begin{tabular}{|c|l|l|}
\hline 
SNe Type & 2002cx group & 2005E group\tabularnewline
\hline
\hline 
Spectral type & Ia: no H and He lines & Ib: no H and Si lines\tabularnewline
 & Si lines & He lines\tabularnewline
\hline 
Absolute magnitude & $-17$ & $-15$\tabularnewline
\hline 
Ejecta velocity & Low & High\tabularnewline
\hline 
Early  & Resemblance to SN 1991T  & Resemblance to regular type Ib SNe\tabularnewline
spectra & Strong Fe-group elements & No Fe-group elements\tabularnewline
\hline 
Late & Fe-group elements  & Intermediate elements (Ca, O) \tabularnewline
spectra & Intermediate elements & No Fe-group elements\tabularnewline
 &  & large ratio of Ca to O, S and Ni \tabularnewline
\hline 
Decline Rate & Slow & Fast\tabularnewline
\hline 
Host galaxies & Late type & Early type\tabularnewline
\hline
\end{tabular}

\caption{\label{t:diffs}  \footnotesize Brief summary of the differences 
between the typical observed properties
of SN 2002cx-like and SN 2005E-like SNe.}

\end{scriptsize}
\end{table}

\bigskip
 \noindent{\bf(11) Additional calcium-rich faint type Ib/c SNe}

In addition to SN 2005E, several other objects were reported as 
possible members of this class of calcium-rich SNe. 
We have verified these reports by reinspection of the 
spectra, rejecting unconvincing cases, 
and list all verified Ca-rich events in Table 3. 
We also note that no radio signature has been found for any 
of these Ca-rich SNe (1--2 would be expected 
for a sample of 8 SNe~Ib \cite{sod+06b}). 

The rate of calcium-rich, faint, SNe~Ib/c can be estimated 
since SN 2005E 
was discovered as part of the Lick Observatory Supernova Search (LOSS)\cite{fil+01}. 
This survey is a volume-limited search,
with high sensitivity within 60 Mpc for both SNe Ia and 
faint Ca-rich objects such as SN 2005E. 
LOSS found 2.3 calcium-rich objects (after correction for incompleteness) 
and 31.0 SNe~Ia in this volume, from which we infer 
the rate of calcium-rich SNe to be 7\% $\pm$ 5\% of the total 
SN~Ia rate.

\clearpage
\begin{table}
\begin{tabular}{|c|c|c|c|c}
\hline 
SN & Absolute $B$-band & Absolute $B$-band     & Host galaxy & Host-galaxy type\tabularnewline
   & peak magnitude  & discovery magnitude &             &                 \tabularnewline
\hline
\hline 
2000ds & ?      & $-13.32^{*}$ & NGC 2768 & E/S0\tabularnewline
\hline 
2001co & $-15.09^{*}$ & $-14.77^{*}$  & NGC 5559 & Sb\tabularnewline
\hline 
2003H  & ?      & $-13.43^{*}$ & NGC 2207 & Galaxy pair\tabularnewline
\hline 
2003dg & ?      & $-15.03^{*}$ & UGC 6934 & Scd\tabularnewline
\hline 
2003dr & $-14.04^{*}$ & $-13.8^{*}$  & NGC 5714 & Sc\tabularnewline
\hline 
2005cz & ? & ?  & NGC 4589 & E\tabularnewline
\hline
2005E  & $-14.8$  & -14.7  & NGC 1132 & S0/Sa\tabularnewline
\hline 
2007ke & $-15.45^{*}$ & $-14.7^{*}$  & NGC 1129 & E/S0\tabularnewline
\hline
\end{tabular} 
\caption{\label{t:sne} { \footnotesize The sample of calcium-rich 
SNe. Discovery of these SNe is reported in 
Refs. \cite{puc+01}(2000ds), ~\cite{aza+01}(2001co),~ \cite{gra+03}(2003H),~ 
\cite{pug+03}(2003dg),~ \cite{puc+03}(2003dr),~ \cite{gra+05}(2005E),~ 
\cite{chu+07}(2007ke), and \cite{kaw+09}(2005cz).}
\newline \footnotesize
$^{*}B$ magnitude unavailable; unfiltered used, corrected to $B$ using 
the measured colors of SN 2005E at peak.} 
\end{table}

A full analysis of the photometry and spectroscopy of our extended sample 
of faint Ca-rich SNe will be presented in a forthcoming publication.

Of the total sample of SNe~Ib/c reported to reside in early-type 
(elliptical or S0) 
galaxies\cite{van+05,hak+08}, SN 2000ds and SN 2005cz are
Ca-rich, faint SNe Ib. The host galaxies of all other putative SNe~Ib/c 
in early-type 
hosts (SNe 2002jj, 2002hz, 2003ih, and 2006ab) were reclassified 
in Ref. \cite{hak+08} to be late-type galaxies. 
Thus, {\it all} confirmed SNe~Ib/c in early-type hosts belong to the faint, 
Ca-rich type Ib class. In Fig. 3 we show the host-galaxy 
distribution for the SN 2005E-like events and compare it 
with the distributions of other types of SNe. The hosts of SN 2005E-like 
events are clearly different from those of SNe~Ib/c and SNe~II (resulting 
from core collapse of a massive, short-lived star\cite{sma09}) 
and show strong preference toward early-type host galaxies, 
and hence older and lower-mass progenitors.   

\newpage   
\bigskip
\noindent{\bf(12) Nucleosynthetic simulations and derived abundances}

Given the unique nucleosynthetic products 
we observed in spectra of SN 2005E, we ran nucleosynthesis single-zone 
simulations\cite{arn96} in order to investigate
possible conditions that may lead to such signatures. 

Our single-zone simulations made use of the JINA version of Reaclib 
(April 2009) for the rates, a 203-nucleus network, and the 
``explosive nucleosynthesis'' procedure\cite{arn96}.
 We investigated 
various initial compositions of He, C, and O, at a density of 
$10^6$ g cm$^{-3}$, and scanned the temperature
range (2--4.1) $\times 10^9$ K. Our main findings are
illustrated in Fig. S5 and Table 4. In Fig. S4, we scale 
the mass fractions to fit the observed 
calcium abundance, and plot the observed limits on sulfur and $^{56}$Ni.
We also present the amount of $^{44}$Ti produced. 
A 0.6/0.4 mix of He/O {\bf (a)} works well in reproducing our data only up to 
$T_9=3.5$ where nickel becomes overabundant, and certainly fails at 
higher $T$, where S is 
also a problem. In the allowed phase space ($T_9<3.5$), $^{44}$Ti is 
0.1--0.3 times as abundant as Ca. Increasing the amount of helium {\bf (b)},
permits only low temperatures ($T_9=2.2$); in this case radioactive titanium
and calcium are almost equally produced. Replacing oxygen by pure carbon
{\bf (c)} does not work for nearly equal ratios, as we either get too much 
S or too much Ni, or both. Decreasing the amount of carbon {\bf (d)}
again works only for the lowest temperatures, $T_9=2.2$, leading to
significant production of $^{44}$Ti.  

We also note that the unique nucleosynthetic
signature of SN 2005E, with evidence for He, O, and Ca, but lacking 
Si, S, and Fe/Ni differs physically (as well as in  
spectral appearance) from the Fe/Ni and Si-rich
ejecta of other types of SNe (see Table 1). 
Fe/Ni, S, and Si are typically oxygen-burning products 
(common in both SNe~Ia and core-collapse events), while the Ca-rich/S-poor 
SN 2005E is dominated by helium-burning products, a unique composition 
which is evidence for an underlying physical process not previously seen in other 
events. Such abundances are also not expected from theoretical 
predictions for either core collapse SNe or thermonuclear explosions of 
Chandrasekhar-mass WDs\cite{nom+97a,nom+97b}. 

We conclude that helium-rich models with some C/O contamination
can recover our essential findings. A prediction is that
a substantial amount of $^{44}$Ti (at least 1/10 of Ca, and perhaps
a comparable amount) will be synthesized.
In some cases $^{48}$Cr production exceeds that 
of $^{56}$Ni; its decay may therefore power the light
curve\cite{bil+07,she+09} (via $^{48}$V decay with a 16 day half-life time).
 A realistic model in which
the burning shock traverses layers of varying temperature, density,
and composition will be investigated in a forthcoming publication. 
 
\clearpage
\centerline{\includegraphics[width=4.5in,angle=0]{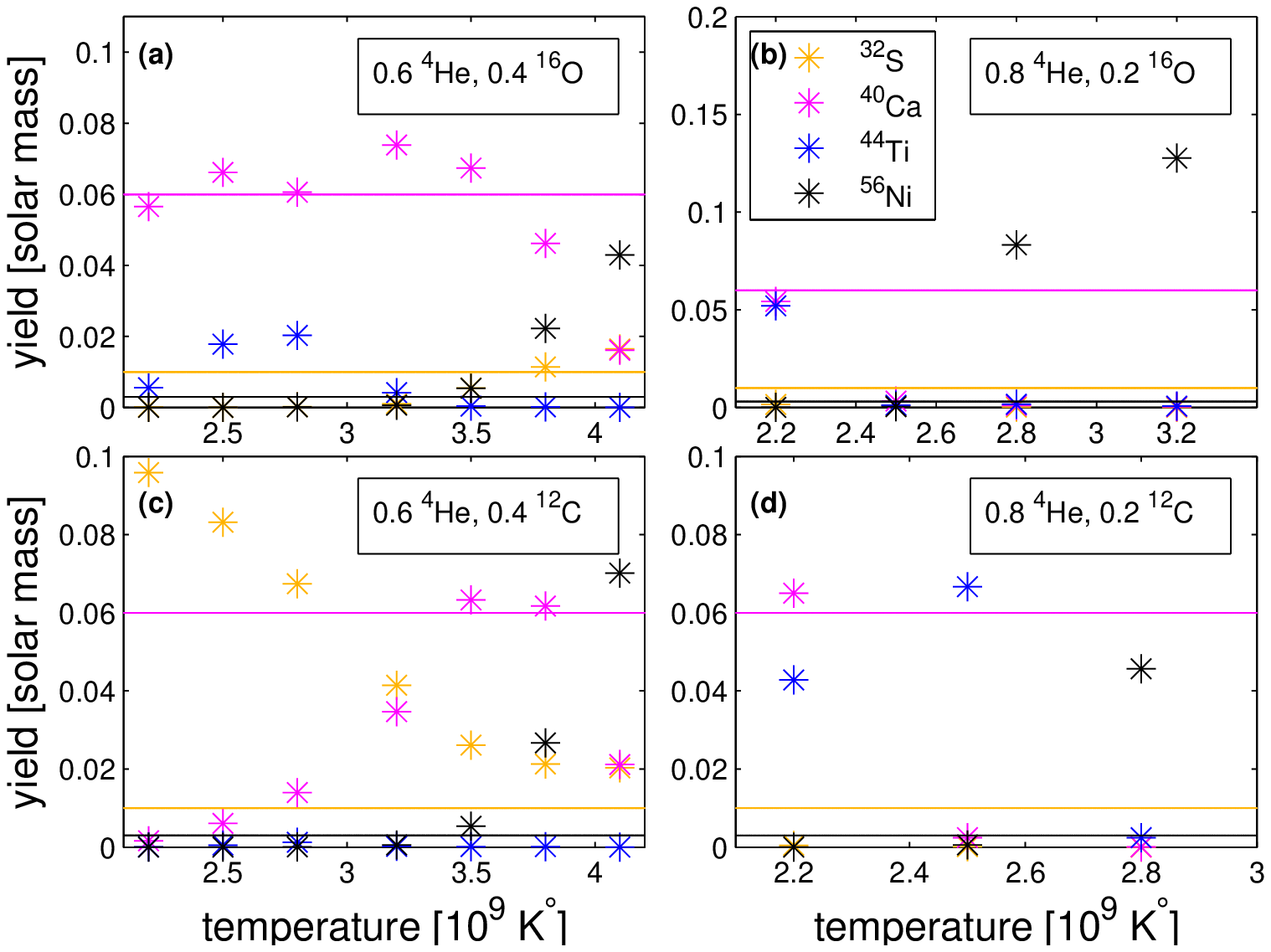}}
\bigskip 
\noindent{\bf Figure S5}:\\
\begin{scriptsize}
  Nucleosynthetic products of the burning of He-rich mixtures
with carbon and oxygen (fractions marked in each panel),
at various temperatures and an initial density of $10^6$ g cm$^{-3}$.
The fractions are scaled to best match the observed calcium 
mass (magenta horizontal line) and compared with the observed limit
on sulfur (ocher line) and measurement of radioactive $^{56}$Ni
(black line).   
\end{scriptsize}
\clearpage
\begin{table}
\begin{scriptsize}

\begin{tabular}{|c|c|c|c|c|c|c|c|c|}
\hline 
\multicolumn{3}{|c|}{Initial Conditions} & \multicolumn{6}{c|}{Final Values}\tabularnewline
\hline 
 $T_{9}$ & $^{4}$He & $^{12}$C & $^{32}$S & $^{40}$Ca & $^{44}$Ti & $^{48}$Cr & $^{52}$Fe & $^{56}$Ni\tabularnewline
\hline
\hline 
$3.4$ & $8.00\times10^{-1}$ & $1.86\times10^{-1}$ & $6.18\times10^{-5}$ & $1.97\times10^{-4}$ & $2.68\times10^{-4}$ & $1.34\times10^{-4}$ & $2.82\times10^{-4}$ & $8.08\times10^{-1}$\tabularnewline
\hline 
$3.4$ & $7.50\times10^{-1}$ & $2.36\times10^{-1}$ & $2.80\times10^{-4}$ & $1.30\times10^{-1}$ & $5.03\times10^{-3}$ & $9.03\times10^{-3}$ & $2.29\times10^{-1}$ & $5.82\times10^{-1}$\tabularnewline
\hline 
$3.4$ & $7.00\times10^{-1}$ & $2.86\times10^{-1}$ & $2.49\times10^{-3}$ & $6.10\times10^{-1}$ & $1.60\times10^{-1}$ & $3.04\times10^{-2}$ & $1.65\times10^{-1}$ & $1.18\times10^{-1}$\tabularnewline
\hline 
$3.4$ & $6.50\times10^{-1}$ & $3.36\times10^{-1}$ & $1.12\times10^{-1}$ & $5.61\times10^{-1}$ & $2.63\times10^{-3}$ & $1.84\times10^{-2}$ & $2.00\times10^{-2}$ & $2.78\times10^{-2}$\tabularnewline
\hline 
$3.4$ & $6.00\times10^{-1}$ & $3.86\times10^{-1}$ & $2.09\times10^{-1}$ & $3.74\times10^{-1}$ & $1.13\times10^{-3}$ & $9.84\times10^{-3}$ & $8.01\times10^{-3}$ & $1.97\times10^{-2}$\tabularnewline
\hline
\noalign{\vskip\doublerulesep}
\hline 
$3.2$ & $8.00\times10^{-1}$ & $1.86\times10^{-1}$ & $5.25\times10^{-5}$ & $1.74\times10^{-4}$ & $2.36\times10^{-4}$ & $1.18\times10^{-4}$ & $3.15\times10^{-4}$ & $8.29\times10^{-1}$\tabularnewline
\hline 
$3.2$ & $7.50\times10^{-1}$ & $2.36\times10^{-1}$ & $1.69\times10^{-6}$ & $1.24\times10^{-1}$ & $3.94\times10^{-1}$ & $6.66\times10^{-2}$ & $4.33\times10^{-1}$ & $2.50\times10^{-1}$\tabularnewline
\hline 
$3.2$ & $7.00\times10^{-1}$ & $2.86\times10^{-1}$ & $4.32\times10^{-4}$ & $6.55\times10^{-1}$ & $1.14\times10^{-1}$ & $8.00\times10^{-2}$ & $7.49\times10^{-2}$ & $1.95\times10^{-2}$\tabularnewline
\hline 
$3.2$ & $6.80\times10^{-1}$ & $2.86\times10^{-1}$ & $2.73\times10^{-2}$ & $6.76\times10^{-1}$ & $2.43\times10^{-2}$ & $2.79\times10^{-2}$ & $1.37\times10^{-2}$ & $9.83\times10^{-3}$\tabularnewline
\hline 
$3.2$ & $6.50\times10^{-1}$ & $3.36\times10^{-1}$ & $1.63\times10^{-1}$ & $4.18\times10^{-1}$ & $4.75\times10^{-3}$ & $1.07\times10^{-2}$ & $4.35\times10^{-3}$ & $6.86\times10^{-3}$\tabularnewline
\hline 
$3.2$ & $6.00\times10^{-1}$ & $3.86\times10^{-1}$ & $2.90\times10^{-1}$ & $2.43\times10^{-1}$ & $1.56\times10^{-3}$ & $5.29\times10^{-3}$ & $1.95\times10^{-3}$ & $4.02\times10^{-3}$\tabularnewline
\hline
\noalign{\vskip\doublerulesep}
\hline 
$3.0$ & $8.00\times10^{-1}$ & $1.86\times10^{-1}$ & $6.06\times10^{-5}$ & $2.15\times10^{-4}$ & $3.26\times10^{-4}$ & $2.38\times10^{-4}$ & $8.69\times10^{-2}$ & $7.50\times10^{-1}$\tabularnewline
\hline 
$3.0$ & $7.50\times10^{-1}$ & $2.36\times10^{-1}$ & $3.85\times10^{-9}$ & $2.96\times10^{-2}$ & $1.73\times10^{-1}$ & $1.78\times10^{-1}$ & $4.53\times10^{-1}$ & $1.24\times10^{-1}$\tabularnewline
\hline 
$3.0$ & $7.00\times10^{-1}$ & $2.86\times10^{-1}$ & $1.06\times10^{-3}$ & $5.72\times10^{-1}$ & $2.71\times10^{-1}$ & $4.41\times10^{-2}$ & $9.32\times10^{-3}$ & $4.16\times10^{-3}$\tabularnewline
\hline 
$3.0$ & $6.50\times10^{-1}$ & $3.36\times10^{-1}$ & $2.40\times10^{-1}$ & $2.88\times10^{-1}$ & $2.91\times10^{-2}$ & $3.77\times10^{-3}$ & $7.91\times10^{-4}$ & $2.34\times10^{-3}$\tabularnewline
\hline 
$3.0$ & $6.00\times10^{-1}$ & $3.86\times10^{-1}$ & $3.85\times10^{-1}$ & $1.53\times10^{-1}$ & $1.14\times10^{-1}$ & $1.51\times10^{-3}$ & $2.47\times10^{-4}$ & $1.44\times10^{-3}$\tabularnewline
\hline
\end{tabular}
\end{scriptsize}

\caption{\label{t:nucleosynthetic} \footnotesize
   The abundances (by mass) of several important nuclei
from post-shock burning in a helium-burning layer. A 203-nucleus 
network was used, with reaction rates taken from an April 2009 download 
of Reaclib from JINA. The initial composition was $^4$He, $^{12}$C (see table 
for initial values; temperature $T_9$ given in $10^{9}$ K), and 0.014 of $^{22}$Ne by mass. The values of $^4$He and 
$^{12}$C correspond to different degrees of hydrostatic helium burning prior 
to explosion. Low values for S/Ca occur only for high $^4$He abundance. 
For such a low S/Ca ratio, the temperature must be $T < 3.2 \times 10^{9}$ K 
to avoid overproduction 
of $^{56}$Ni. Production of $^{48}$Cr (and $^{52}$Fe) may exceed that 
of $^{56}$Ni, so 
that $^{48}$V decay (half-life of 16 days) may help power the light curve\cite{bil+07}. 
$^{44}$Ti can be copiously produced under these conditions, so that such
events may provide a solution to the problem of low emission from $^{44}$Ti 
decay in most nearby SN remnants\cite{tim+96,the+06}. 
The significant abundance of $^{12}$C prior to the explosive burning may allow 
unburned (lower density) regions to explain the relatively large abundance 
of $^{12}$C  observed. To convert the abundance entries to solar mass units, 
multiply by the assumed mass of ejecta.}

\end{table}

\newpage

\clearpage



\begin{thebibliography}{100}

\bibitem[1]{fil97}
{Filippenko}, A.~V. {Optical spectra of supernovae}.
\newblock {\it Ann. Rev. Astron. Astrophys.} {\bf 35}, 309--355 (1997).

\bibitem[2]{maz+07}
{Mazzali}, P.~A. {\it et al.} {A common explosion mechanism for type Ia
  supernovae}.
\newblock {\it Science} {\bf 315}, 825--828 (2007).

\bibitem[3]{fil+92}
{Filippenko}, A.~V. {\it et al.} {The subluminous, spectroscopically peculiar
  type IA supernova 1991bg in the elliptical galaxy NGC 4374}.
\newblock {\it Astron. J.} {\bf 104}, 1543--1556 (1992).

\bibitem[4]{li+03}
{Li}, W. {\it et al.} {SN 2002cx: The most peculiar known type Ia supernova}.
\newblock {\it Pub. Astron. Soc. Pac.} {\bf 115}, 453--473 (2003).

\bibitem[5]{fil+03}
{Filippenko}, A.~V. {\it et al.} {Supernovae 2001co, 2003H, 2003dg, and
  2003dr}.
\newblock {\it IAU Circular} {\bf 8159}, 2 (2003).

\bibitem[6]{pra+03}
{Prada}, F. {\it et al.} {Observing the dark matter density profile of isolated
  galaxies}.
\newblock {\it Astrophys. J.} {\bf 598}, 260--271 (2003).

\bibitem[7]{ken98}
{Kennicutt}, Jr., R.~C. {Star formation in galaxies along the Hubble sequence}.
\newblock {\it Ann. Rev. Astron. Astrophys.} {\bf 36}, 189--232 (1998).

\bibitem[8]{kas+08}
{Kasliwal}, M.~M. {\it et al.} {SN 2007ax: An extremely faint type Ia
  supernova}.
\newblock {\it Astrophys. J. Lett.} {\bf 683}, L29--L32 (2008).

\bibitem[9]{fol+09}
{Foley}, R.~J. {\it et al.} {SN 2008ha: An extremely low luminosity and
  exceptionally low energy supernova}.
\newblock {\it Astron. J.} {\bf 138}, 376--391 (2009).

\bibitem[10]{val+09}
{Valenti}, S. {\it et al.} {A low-energy core-collapse supernova without a
  hydrogen envelope}.
\newblock {\it Nature} {\bf 459}, 674--677 (2009).

\bibitem[11]{kaw+09}
{Kawabata}, K.~S. {\it et al.} {A unique core-collapse supernova in an
  elliptical galaxy}.
\newblock {\it ArXiv} {\bf 0906.2811},  (2009).

\bibitem[12]{mau+05}
{Maund}, J.~R. \& {Smartt}, S.~J. {Hubble Space Telescope imaging of the
  progenitor sites of six nearby core-collapse supernovae}.
\newblock {\it Mon. Not. R. Astron. Soc.} {\bf 360}, 288--304 (2005).

\bibitem[13]{van+05}
{van den Bergh}, S., {Li}, W.  \& {Filippenko}, A.~V. {Classifications of the
  host galaxies of supernovae, set III}.
\newblock {\it Pub. Astron. Soc. Pac.} {\bf 117}, 773--782 (2005).

\bibitem[14]{hak+08}
{Hakobyan}, A.~A. {\it et al.} {Early-type galaxies with core collapse
  supernovae}.
\newblock {\it Astron. Astrophys.} {\bf 488}, 523--531 (2008).

\bibitem[15]{woo+86}
{Woosley}, S.~E., {Taam}, R.~E.  \& {Weaver}, T.~A. {Models for type I
  supernova. I - Detonations in white dwarfs}.
\newblock {\it Astrophys. J.} {\bf 301}, 601--623 (1986).

\bibitem[16]{woo+94}
{Woosley}, S.~E. \& {Weaver}, T.~A. {Sub-Chandrasekhar mass models for type Ia
  supernovae}.
\newblock {\it Astrophys. J.} {\bf 423}, 371--379 (1994).

\bibitem[17]{nom+91}
{Nomoto}, K. \& {Kondo}, Y. {Conditions for accretion-induced collapse of white
  dwarfs}.
\newblock {\it Astrophys. J. Lett.} {\bf 367}, L19--L22 (1991).

\bibitem[18]{met+09}
{Metzger}, B.~D., {Piro}, A.~L.  \& {Quataert}, E. {Nickel-rich outflows from
  accretion disks formed by the accretion-induced collapse of white dwarfs}.
\newblock {\it Mon. Not. R. Astron. Soc.} {\bf 396}, 1659 (2009).

\bibitem[19]{bil+07}
{Bildsten}, L. {\it et al.} {Faint thermonuclear supernovae from AM Canum
  Venaticorum binaries}.
\newblock {\it Astrophys. J. Lett.} {\bf 662}, L95--L98 (2007).

\bibitem[20]{liv+95}
{Livne}, E. \& {Arnett}, D. {Explosions of sub--Chandrasekhar mass white dwarfs
  in two dimensions}.
\newblock {\it Astrophys. J.} {\bf 452}, 62--74 (1995).

\bibitem[21]{ibe+87}
{Iben}, I.~J. {\it et al.} {On interacting helium star-white dwarf pairs as
  supernova precursors}.
\newblock {\it Astrophys. J.} {\bf 317}, 717--723 (1987).

\bibitem[22]{woo+73}
{Woosley}, S.~E., {Arnett}, W.~D.  \& {Clayton}, D.~D. {The explosive burning
  of oxygen and silicon}.
\newblock {\it Astrophys. J. Supp. Ser.} {\bf 26}, 231--312 (1973).

\bibitem[23]{tim+96}
{Timmes}, F.~X. {\it et al.} {The production of 44Ti and 60Co in supernovae}.
\newblock {\it Astrophys. J.} {\bf 464}, 332--341 (1996).

\bibitem[24]{lai+09}
{Lai}, D.~K. {\it et al.} {A unique star in the outer halo of the Milky Way}.
\newblock {\it Astrophys. J. Lett.} {\bf 697}, L63--L67 (2009).

\bibitem[25]{dep+07}
{de Plaa}, J. {\it et al.} {Constraining supernova models using the hot gas in
  clusters of galaxies}.
\newblock {\it Astron. Astrophys.} {\bf 465}, 345--355 (2007).

\bibitem[26]{the+06}
{The}, L.-S. {\it et al.} {Are $^{44}$Ti-producing supernovae exceptional?}
\newblock {\it Astron. Astrophys.} {\bf 450}, 1037--1050 (2006).

\bibitem[27]{cha+93}
{Chan}, K.-W. \& {Lingenfelter}, R.~E. {Positrons from supernovae}.
\newblock {\it Astrophys. J.} {\bf 405}, 614--636 (1993).

\bibitem[28]{kno+05}
{Kn{\"o}dlseder}, J. {\it et al.} {The all-sky distribution of 511 keV
  electron-positron annihilation emission}.
\newblock {\it Astron. Astrophys.} {\bf 441}, 513--532 (2005).

\bibitem[29]{fil+01}
{Filippenko}, A.~V. {\it et al.} {The Lick Observatory Supernova Search with
  the Katzman Automatic Imaging Telescope}.
\newblock {\it Astronomical Society of the Pacific Conference Series} {\bf
  246}, 121--130 (2001).

\bibitem[30]{oke+95}
{Oke}, J.~B. {\it et al.} {The Keck low-resolution imaging spectrometer}.
\newblock {\it Pub. Astron. Soc. Pac.} {\bf 107}, 375--385 (1995).

\bibitem[31]{gal+07}
{Gal-Yam}, A. {\it et al.} {Preliminary Results from the Caltech Core-Collapse
  Project (CCCP)}.
\newblock {\it American Institute of Physics Conference Series} {\bf 924},
  297--303 (2007).

\bibitem[32]{oke+82}
{Oke}, J.~B. \& {Gunn}, J.~E. {An efficient low resolution and moderate
  resolution spectrograph for the Hale telescope}.
\newblock {\it Pub. Astron. Soc. Pac.} {\bf 94}, 586--594 (1982).

\bibitem[33]{maz+97}
{Mazzali}, P.~A. {\it et al.} {The properties of the peculiar type Ia supernova
  1991bg - II. The amount of \^{}56Ni and the total ejecta mass determined from
  spectrum synthesis and energetics considerations}.
\newblock {\it Mon. Not. R. Astron. Soc.} {\bf 284}, 151--171 (1997).

\bibitem[34]{maz+04}
{Mazzali}, P.~A. {\it et al.} {Properties of two hypernovae entering the
  nebular phase: SN 1997ef and SN 1997dq}.
\newblock {\it Astrophys. J.} {\bf 614}, 858--863 (2004).

\bibitem[35]{how+05}
{Howell}, D.~A. {\it et al.} {Gemini spectroscopy of supernovae from the
  supernova legacy survey: Improving high-redshift supernova selection and
  classification}.
\newblock {\it Astrophys. J.} {\bf 634}, 1190--1201 (2005).

\bibitem[36]{chu+08}
{Chu}, Y.-H. \& {Gruendl}, R.~A. in {\it Massive star formation: Observations
  confront theory} (eds {Beuther}, H., {Linz}, H.  \& {Henning}, T.)  415--422
  (ASP, 2008).

\bibitem[37]{sch+08}
{Schilbach}, E. \& {R{\"o}ser}, S. {On the origin of field O-type stars}.
\newblock {\it Astron. Astrophys.} {\bf 489}, 105--114 (2008).

\bibitem[38]{and+08}
{Anderson}, J.~P. \& {James}, P.~A. {Constraints on core-collapse supernova
  progenitors from correlations with H{$\alpha$} emission}.
\newblock {\it Mon. Not. R. Astron. Soc.} {\bf 390}, 1527--1538 (2008).

\bibitem[39]{dys+83}
{Dyson}, J.~E. \& {Hartquist}, T.~W. {On the structure of intermediate- and
  high-velocity clouds}.
\newblock {\it Mon. Not. R. Astron. Soc.} {\bf 203}, 1233--1238 (1983).

\bibitem[40]{chr+97}
{Christodoulou}, D.~M., {Tohline}, J.~E.  \& {Keenan}, F.~P. {Star-forming
  processes far from the galactic disk: inoperative or indolent where
  operative}.
\newblock {\it Astrophys. J.} {\bf 486}, 810--817 (1997).

\bibitem[41]{mar+99}
{Martos}, M. {\it et al.} {Spiral density wave shock-induced star formation at
  high galactic latitudes}.
\newblock {\it Astrophys. J. Lett.} {\bf 526}, L89--L92 (1999).

\bibitem[42]{her+96}
{Heraudeau}, P. \& {Simien}, F. {Optical and I-band surface photometry of
  spiral galaxies. I. The data.}
\newblock {\it AAPS} {\bf 118}, 111--155 (1996).

\bibitem[43]{ken+83}
{Kennicutt}, R.~C., J. \& {Kent}, S.~M. {A survey of H-alpha emission in normal
  galaxies}.
\newblock {\it Astron. J.} {\bf 88}, 1094--1107 (1983).

\bibitem[44]{spr+05}
{Springob}, C.~M. {\it et al.} {A Digital Archive of H I 21 Centimeter Line
  Spectra of Optically Targeted Galaxies}.
\newblock {\it Astrophys. J. Supp. Ser.} {\bf 160}, 149--162 (2005).

\bibitem[45]{wel+03}
{Welch}, G.~A. \& {Sage}, L.~J. {The Cool Interstellar Medium in S0 Galaxies.
  I. A Survey of Molecular Gas}.
\newblock {\it Astrophys. J.} {\bf 584}, 260--277 (2003).

\bibitem[46]{sod07}
{Soderberg}, A.~M. {The radio properties of type Ibc supernovae}.
\newblock {\it American Institute of Physics Conference Series} {\bf 937},
  492--499 (2007).

\bibitem[47]{ham+02}
{Hamuy}, M. {\it et al.} {Optical and infrared spectroscopy of SN 1999ee and SN
  1999ex}.
\newblock {\it Astron. J.} {\bf 124}, 417--429 (2002).

\bibitem[48]{fil+86}
{Filippenko}, A.~V. \& {Sargent}, W.~L.~W. {The unique supernova (1985f) in NGC
  4618}.
\newblock {\it Astron. J.} {\bf 91}, 691--696 (1986).

\bibitem[49]{gas+86}
{Gaskell}, C.~M. {\it et al.} {Type Ib supernovae 1983n and 1985f - Oxygen-rich
  late time spectra}.
\newblock {\it Astrophys. J. Lett.} {\bf 306}, L77--L80 (1986).

\bibitem[50]{nad03}
{Nadyozhin}, D.~K. {Explosion energies, nickel masses and distances of type II
  plateau supernovae}.
\newblock {\it Mon. Not. R. Astron. Soc.} {\bf 346}, 97--104 (2003).

\bibitem[51]{maz+00}
{Mazzali}, P.~A., {Iwamoto}, K.  \& {Nomoto}, K. {A spectroscopic analysis of
  the energetic type Ic Hypernova SN 1997EF}.
\newblock {\it Astrophys. J.} {\bf 545}, 407--419 (2000).

\bibitem[52]{tom+05}
{Tominaga}, N. {\it et al.} {The unique type Ib supernova 2005bf: A WN star
  explosion model for peculiar light curves and spectra}.
\newblock {\it Astrophys. J. Lett.} {\bf 633}, L97--L100 (2005).

\bibitem[53]{zam05}
{Zampieri}, L. {Physical properties of type II supernovae and their
  progenitors}.
\newblock {\it Astronomical Society of the Pacific Conference Series} {\bf
  342}, 358--365 (2005).

\bibitem[54]{li06}
{Li}, L.-X. {Correlation between the peak spectral energy of gamma-ray bursts
  and the peak luminosity of the underlying supernovae: implication for the
  nature of the gamma-ray burst-supernova connection}.
\newblock {\it Mon. Not. R. Astron. Soc.} {\bf 372}, 1357--1365 (2006).

\bibitem[55]{sau+06}
{Sauer}, D.~N. {\it et al.} {The properties of the `standard' Type Ic supernova
  1994I from spectral models}.
\newblock {\it Mon. Not. R. Astron. Soc.} {\bf 369}, 1939--1948 (2006).

\bibitem[56]{maz+07b}
{Mazzali}, P.~A. {\it et al.} {The aspherical properties of the energetic type
  Ic SN 2002ap as inferred from its nebular spectra}.
\newblock {\it Astrophys. J.} {\bf 670}, 592--599 (2007).

\bibitem[57]{utr+07}
{Utrobin}, V.~P., {Chugai}, N.~N.  \& {Pastorello}, A. {Ejecta and progenitor
  of the low-luminosity type IIP supernova 2003Z}.
\newblock {\it Astron. Astrophys.} {\bf 475}, 973--979 (2007).

\bibitem[58]{sod+06a}
{Soderberg}, A.~M. {\it et al.} {An HST study of the supernovae accompanying
  GRB 040924 and GRB 041006}.
\newblock {\it Astrophys. J.} {\bf 636}, 391--399 (2006).

\bibitem[59]{pas+04}
{Pastorello}, A. {\it et al.} {Low-luminosity Type II supernovae: spectroscopic
  and photometric evolution}.
\newblock {\it Mon. Not. R. Astron. Soc.} {\bf 347}, 74--94 (2004).

\bibitem[60]{arn82}
{Arnett}, W.~D. {Type I supernovae. I - Analytic solutions for the early part
  of the light curve}.
\newblock {\it Astrophys. J.} {\bf 253}, 785--797 (1982).

\bibitem[61]{ste+05}
{Stehle}, M. {\it et al.} {Abundance stratification in Type Ia supernovae - I.
  The case of SN 2002bo}.
\newblock {\it Mon. Not. R. Astron. Soc.} {\bf 360}, 1231--1243 (2005).

\bibitem[62]{mod+09}
{Modjaz}, M. {\it et al.} {From shock breakout to peak and beyond: Extensive
  panchromatic observations of the aspherical type Ib supernova 2008D
  associated with Swift X-ray transient 080109}.
\newblock {\it Astrophys. J.} {\bf 702}, 226 (2009).

\bibitem[63]{cen+06}
{Cenko}, S.~B. {\it et al.} {The Automated Palomar 60 Inch Telescope}.
\newblock {\it Pub. Astron. Soc. Pac.} {\bf 118}, 1396--1406 (2006).

\bibitem[64]{ned}
NED {\it Nasa/ipac extragalactic database} http://nedwww.ipac.caltech.edu/.

\bibitem[65]{tan+07}
{Tanaka}, M., {Maeda}, K., {Mazzali}, P.~A.  \& {Nomoto}, K. {Multidimensional
  Simulations for Early-Phase Spectra of Aspherical Hypernovae: SN 1998bw and
  Off-Axis Hypernovae}.
\newblock {\it Astrophys. J. Lett.} {\bf 668}, L19--L22 (2007).

\bibitem[66]{maz+07c}
{Mazzali}, P.~A. {\it et al.} {Keck and European Southern Observatory Very
  Large Telescope View of the Symmetry of the Ejecta of the XRF/SN 2006aj}.
\newblock {\it Astrophys. J.} {\bf 661}, 892--898 (2007).

\bibitem[67]{sau+08}
{Sauer}, D.~N. \& {Mazzali}, P.~A. {Interpretation of observed type Ia
  supernova spectra with radiative transfer models}.
\newblock {\it New Astron. Rev.} {\bf 52}, 370--372 (2008).

\bibitem[68]{sah+08}
{Sahu}, D.~K. {\it et al.} {The evolution of the peculiar type Ia supernova SN
  2005hk over 400 days}.
\newblock {\it Astrophys. J.} {\bf 680}, 580--592 (2008).

\bibitem[69]{bla61}
Blaauw, A. {On the origin of the O- and B-type stars with high velocities (the
  "run-away" stars), and some related problems}.
\newblock {\it Bulletin of the Astronomical Institutes of the Netherlands} {\bf
  15}, 265--290 (1961).

\bibitem[70]{hoo+01}
{Hoogerwerf}, R., {de Bruijne}, J.~H.~J.  \& {de Zeeuw}, P.~T. {On the origin
  of the O and B-type stars with high velocities. II. Runaway stars and pulsars
  ejected from the nearby young stellar groups}.
\newblock {\it Astron. Astrophys.} {\bf 365}, 49--77 (2001).

\bibitem[71]{hil88}
{Hills}, J.~G. {Hyper-velocity and tidal stars from binaries disrupted by a
  massive Galactic black hole}.
\newblock {\it Nature} {\bf 331}, 687--689 (1988).

\bibitem[72]{bro+05}
{Brown}, W.~R. {\it et al.} {Discovery of an unbound hypervelocity star in the
  Milky Way halo}.
\newblock {\it Astrophys. J. Lett.} {\bf 622}, L33--L36 (2005).

\bibitem[73]{ede+06}
{Edelmann}, H. {\it et al.} {HE 0437-5439: An unbound hypervelocity
  main-sequence B-Type star}.
\newblock {\it Astrophys. J. Lett.} {\bf 634}, L181--L184 (2005).

\bibitem[74]{hir+05}
{Hirsch}, H.~A. {\it et al.} {US 708 - an unbound hyper-velocity subluminous O
  star}.
\newblock {\it Astron. Astrophys.} {\bf 444}, L61--L64 (2005).

\bibitem[75]{eis+05}
{Eisenhauer}, F. {\it et al.} {SINFONI in the Galactic center: Young stars and
  infrared flares in the central light-month}.
\newblock {\it Astrophys. J.} {\bf 628}, 246--259 (2005).

\bibitem[76]{ghe+05}
{Ghez}, A.~M. {\it et al.} {Stellar orbits around the Galactic center black
  hole}.
\newblock {\it Astrophys. J.} {\bf 620}, 744--757 (2005).

\bibitem[77]{miy+75}
{Miyamoto}, M. \& {Nagai}, R. {Three-dimensional models for the distribution of
  mass in galaxies}.
\newblock {\it Publ. of the Astronomical Society of Japan} {\bf 27}, 533--543
  (1975).

\bibitem[78]{gor+07}
{Gorgas}, J., {Jablonka}, P.  \& {Goudfrooij}, P. {Stellar population gradients
  in bulges along the Hubble sequence. I. The data}.
\newblock {\it Astron. Astrophys.} {\bf 474}, 1081--1092 (2007).

\bibitem[79]{fer+00}
{Ferrarese}, L. \& {Merritt}, D. {A fundamental relation between supermassive
  black holes and their host galaxies}.
\newblock {\it Astrophys. J. Lett.} {\bf 539}, L9--L12 (2000).

\bibitem[80]{geb+00}
{Gebhardt}, K. {\it et al.} {A relationship between nuclear black hole mass and
  galaxy velocity dispersion}.
\newblock {\it Astrophys. J. Lett.} {\bf 539}, L13--L16 (2000).

\bibitem[81]{sha+06}
{Shankar}, F. {\it et al.} {New relationships between galaxy properties and
  host halo mass, and the role of feedbacks in galaxy formation}.
\newblock {\it Astrophys. J.} {\bf 643}, 14--25 (2006).

\bibitem[82]{leo+91}
{Leonard}, P.~J.~T. {The maximum possible velocity of dynamically ejected
  runaway stars}.
\newblock {\it Astron. J.} {\bf 101}, 562--571 (1991).

\bibitem[83]{leo+90}
{Leonard}, P.~J.~T. \& {Duncan}, M.~J. {Runaway stars from young star clusters
  containing initial binaries. II - A mass spectrum and a binary energy
  spectrum}.
\newblock {\it Astron. J.} {\bf 99}, 608--616 (1990).

\bibitem[84]{hil91}
{Hills}, J.~G. {Computer simulations of encounters between massive black holes
  and binaries}.
\newblock {\it Astron. J.} {\bf 102}, 704--715 (1991).

\bibitem[85]{bro+06c}
{Bromley}, B.~C. {\it et al.} {Hypervelocity stars: Predicting the spectrum of
  ejection velocities}.
\newblock {\it Astrophys. J.} {\bf 653}, 1194--1202 (2006).

\bibitem[86]{abt83}
{Abt}, H.~A. {Normal and abnormal binary frequencies}.
\newblock {\it Ann. Rev. Astron. Astrophys.} {\bf 21}, 343--372 (1983).

\bibitem[87]{mor+91}
{Morrell}, N. \& {Levato}, H. {Spectroscopic binaries in the Orion OB1
  association}.
\newblock {\it Astrophys. J. Supp. Ser.} {\bf 75}, 965--985 (1991).

\bibitem[88]{kob+07}
{Kobulnicky}, H.~A. \& {Fryer}, C.~L. {A new look at the binary characteristics
  of massive stars}.
\newblock {\it Astrophys. J.} {\bf 670}, 747--765 (2007).

\bibitem[89]{per09b}
{Perets}, H.~B. {Runaway and hypervelocity stars in the Galactic halo: Binary
  rejuvenation and triple disruption}.
\newblock {\it Astrophys. J.} {\bf 698}, 1330--1340 (2009).

\bibitem[90]{heb+08}
{Heber}, U. {\it et al.} {The B-type giant HD 271791 in the Galactic halo.
  Linking run-away stars to hyper-velocity stars}.
\newblock {\it Astron. Astrophys.} {\bf 483}, L21--L24 (2008).

\bibitem[91]{yuq+03}
{Yu}, Q. \& {Tremaine}, S. {Ejection of hypervelocity stars by the (binary)
  black hole in the Galactic center}.
\newblock {\it Astrophys. J.} {\bf 599}, 1129--1138 (2003).

\bibitem[92]{per+07}
{Perets}, H.~B., {Hopman}, C.  \& {Alexander}, T. {Massive perturber-driven
  interactions between stars and a massive black hole}.
\newblock {\it Astrophys. J.} {\bf 656}, 709--720 (2007).

\bibitem[93]{per09}
{Perets}, H.~B. {Dynamical and evolutionary constraints on the nature and
  origin of hypervelocity stars}.
\newblock {\it Astrophys. J.} {\bf 690}, 795--801 (2009).

\bibitem[94]{per+09}
{Perets}, H.~B. {\it et al.} {Dynamical evolution of the young stars in the
  Galactic center: N-body simulations of the S-Stars}.
\newblock {\it Astrophys. J.} {\bf 702}, 884--889 (2009).

\bibitem[95]{gou+03}
{Gould}, A. \& {Quillen}, A.~C. {Sagittarius A$^{*}$ companion S0-2: A probe of
  very high mass star formation}.
\newblock {\it Astrophys. J.} {\bf 592}, 935--940 (2003).

\bibitem[96]{gil+09}
{Gillessen}, S. {\it et al.} {Monitoring stellar orbits around the massive
  black hole in the Galactic center}.
\newblock {\it Astrophys. J.} {\bf 692}, 1075--1109 (2009).

\bibitem[97]{jha+06}
{Jha}, S. {\it et al.} {Late-time spectroscopy of SN 2002cx: The prototype of a
  new subclass of type Ia supernovae}.
\newblock {\it Astron. J.} {\bf 132}, 189--196 (2006).

\bibitem[98]{sod+06b}
{Soderberg}, A.~M. {\it et al.} {Late-time radio observations of 68 type Ibc
  supernovae: Strong constraints on off-Axis gamma-ray bursts}.
\newblock {\it Astrophys. J.} {\bf 638}, 930--937 (2006).

\bibitem[99]{puc+01}
{Puckett}, T. \& {Dowdle}, G. {Supernova 2000ds in NGC 2768}.
\newblock {\it IAU circulars} {\bf 7507}, 2 (2000).

\bibitem[100]{aza+01}
{Aazami}, A.~B. \& {Li}, W.~D. {Supernova 2001co in NGC 5559}.
\newblock {\it IAU circulars} {\bf 7643}, 2 (2001).

\bibitem[101]{gra+03}
{Graham}, J. {\it et al.} {Supernovae 2003E, 2003F, 2003G, 2003H}.
\newblock {\it IAU circulars} {\bf 8045}, 1 (2003).

\bibitem[102]{pug+03}
{Pugh}, H. \& {Li}, W. {Supernova 2003dg in UGC 6934}.
\newblock {\it IAU circulars} {\bf 8113}, 2 (2003).

\bibitem[103]{puc+03}
{Puckett}, T. {\it et al.} {Supernovae 2003dm, 2003dn, 2003do, 2003dp, 2003dq,
  2003dr}.
\newblock {\it IAU circulars} {\bf 8117}, 1 (2003).

\bibitem[104]{gra+05}
{Graham}, J. {\it et al.} {Supernovae 2005E, 2005F, 2005G, 2005H, 2005I,
  2005J}.
\newblock {\it IAU circulars} {\bf 8467}, 1 (2005).

\bibitem[105]{chu+07}
{Chu}, J. \& {Li}, W. {Supernova 2007ke in NGC 1129}.
\newblock {\it Central Bureau Electronic Telegrams} {\bf 1084}, 1 (2007).

\bibitem[106]{sma09}
{Smartt}, S.~J. {Progenitors of Core-Collapse Supernovae}.
\newblock {\it Ann. Rev. Astron. Astrophys.} {\bf 47}, 63--106 (2009).

\bibitem[107]{arn96}
{Arnett}, D.
\newblock {\it {Supernovae and nucleosynthesis. an investigation of the history
  of matter, from the Big Bang to the present}}.
\newblock Princeton series in astrophysics, Princeton, NJ: Princeton University
  Press (1996).

\bibitem[108]{nom+97a}
{Nomoto}, K. {\it et al.} {Nucleosynthesis in type Ia supernovae.}
\newblock {\it Nuclear Physics A} {\bf 621}, 467--476 (1997).

\bibitem[109]{nom+97b}
{Nomoto}, K. {\it et al.} {Nucleosynthesis in type II supernovae.}
\newblock {\it Nuclear Physics A} {\bf 616}, 79--90 (1997).

\bibitem[110]{she+09}
{Shen}, K.~J. \& {Bildsten}, L. {Unstable helium shell burning on accreting
  white dwarfs}.
\newblock {\it Astrophys. J.} {\bf 699}, 1365 (2009).

\end{thebibliography}

\clearpage

\end{document}